\documentclass[acmsmall,nonacm]{acmart}
\usepackage{bbold}
\usepackage{tikz}
\usepackage{tikz}
\usepackage{url}
\usepackage{amsmath}
\usepackage{mathtools}
\usepackage{hyperref}
\usepackage{braket}
\usetikzlibrary{positioning}
\usepackage{listings}
\usepackage{tcolorbox}
\usepackage[table]{xcolor}
\usepackage{subcaption}
\usepackage{tabularx}
\usepackage{array}
\newtheorem{definition}{Definition}[section]
\usepackage{multirow}
\usepackage{caption}
\usepackage[capitalize]{cleveref}
\usetikzlibrary{calc}
\usepackage{physics}
\usepackage{booktabs}
\usepackage{adjustbox}
\usepackage{quantikz}
\usepackage{wrapfig}

\usepackage{tcolorbox}
\usepackage{xcolor}
\definecolor{rqred}{RGB}{0,76,153}

\newtcolorbox{rqanswer}[1]{
    width=\linewidth,
    colback=white,
    colframe=rqred,
    colbacktitle=rqred,
    coltitle=white,
    fonttitle=\bfseries,
    title={#1},
    boxrule=1.2pt,
    arc=0pt,
    outer arc=0pt,
    boxsep=2pt,
    left=6pt,
    right=6pt,
    top=4pt,
    bottom=4pt,
    toptitle=3pt,
    bottomtitle=1pt,
    lefttitle=6pt,
    righttitle=6pt
}

\definecolor{graphbox}{HTML}{EAF2FF}
\definecolor{prepbox}{HTML}{EAF7EA}
\definecolor{measbox}{HTML}{FFF3E6}
\definecolor{pythonblue}{HTML}{1F77B4}

\usetikzlibrary{positioning, arrows.meta}
\AtBeginDocument{%
  }

\usepackage[dvipsnames]{xcolor}

\lstdefinestyle{qiskit}{
	language=Python,
	basicstyle=\ttfamily\scriptsize,
	keywordstyle=\color{blue!70!black}\bfseries,
	commentstyle=\color{green!45!black}\itshape,
	stringstyle=\color{red!60!black},
	numbers=left,
	numberstyle=\tiny\color{gray},
	numbersep=5pt,
	frame=single,
	rulecolor=\color{gray!60},
	backgroundcolor=\color{gray!5},
	xleftmargin=1.5em,
	framexleftmargin=1.5em,
	showstringspaces=false,
	keepspaces=true
}

\newcommand{\one}[0]{\mathbb{1}}

\usepackage{marvosym}
\makeatletter
\def\blfootnote{\gdef\@thefnmark{}\@footnotetext}
\def\@mkauthors{%
  \global\setbox\mktitle@bx=\vbox{\hsize=\textwidth
    \noindent\unvbox\mktitle@bx\par\medskip
    \noindent{\@authorfont\paperauthors}\par\smallskip
    {\@affiliationfont\paperaffiliations}\par\medskip}}
\makeatother
\newcommand{\authorinfo}[3]{%
  \texorpdfstring{\textsuperscript{#1}\,\orcidlink{#2}\,%
    \href{mailto:#3}{\raisebox{0.1ex}{\small\Letter}}}{}}

\usepackage{braket}
\usepackage{orcidlink}
\usepackage{pgfplots}
\pgfplotsset{compat=1.18}

\usepgfplotslibrary{statistics}

\definecolor{boxbaseline}{HTML}{2A78D6}
\definecolor{boxlconly}{HTML}{EB6834}
\definecolor{boxlcshuffle}{HTML}{1BAF7A}
\definecolor{boxall}{HTML}{8E5BD1}
\definecolor{boxcnot}{HTML}{E8B43A}

\definecolor{generation}{HTML}{2A78D6}
\definecolor{mr}{HTML}{EB6834}
\definecolor{optimization}{HTML}{E87BA4}
\definecolor{execution}{HTML}{1BAF7A}

\begin{document}

\settopmatter{printacmref=false}
\setcopyright{none}

\title{Metamorphic Testing of Quantum Software based on Stabilizer States}

\newcommand{\paperauthors}{%
  Adri\'an Mar\'in Boyero\authorinfo{1,*}{0000-0002-7531-2418}{Adrian.MarinBoyero@uni-siegen.de},
  Fynn Otto\authorinfo{2,*}{0009-0009-0382-042X}{Fynn.Otto@uni-siegen.de},
  Albert Rico\authorinfo{2}{0000-0001-8211-499X}{Albert.RicoAndres@uni-siegen.de},
  Otfried G\"uhne\authorinfo{2}{0000-0002-6033-0867}{Otfried.Guehne@uni-siegen.de},
  and Malte Lochau\authorinfo{1}{0000-0002-8404-753X}{Malte.Lochau@uni-siegen.de}}
\newcommand{\paperaffiliations}{%
  \noindent\textsuperscript{1}Department Elektrotechnik und Informatik, Naturwissenschaftlich-Technische Fakult\"{a}t, Universit\"{a}t Siegen, H\"olderlinstra\ss e 3, 57076 Siegen, Germany\par
  \noindent\textsuperscript{2}Department Physik, Naturwissenschaftlich-Technische Fakult\"{a}t, Universit\"{a}t Siegen, Walter-Flex-Stra\ss e 3, 57068 Siegen, Germany}

\author{Adri\'an Mar\'in Boyero, Fynn Otto, Albert Rico, Otfried G\"uhne, and Malte Lochau}
\authorsaddresses{}

\begin{abstract}
A major challenge in quantum computing is to systematically check if
a quantum program implements its desired functionality.
For classical programs, testing is widely used; 
each test case comprises a pair of sample input and expected 
output to be compared with the actual output.
To increase the number of test cases,
however, one has to solve the test-oracle problem: 
for each additional test input one has to know the correct output.
Metamorphic testing tackles this problem by utilizing
metamorphic relations between input and output domains 
to automatically derive new test cases from existing ones. 
However, adapting metamorphic testing to the quantum realm is not straightforward, 
since quantum states cannot be directly inspected.

We propose a framework of
quantum metamorphic testing (QMT) in which the metamorphic relations 
incorporate the measurement procedures required to assess quantum outputs.
We apply our QMT framework to testing the preparation of stabilizer states,
which are central to protocols like measurement-based quantum computing or quantum error correction.
We utilize structural properties of stabilizer states to 
systematically derive measurement circuits for new test cases from existing ones.
We implement this framework in Qiskit, and demonstrate that it reduces by an order of magnitude the testing run time compared to testing with randomly generated input states.
We further show effectiveness by revealing synthetic program faults
injected by circuit mutations, as well as a real bug in the Qiskit compiler.
\end{abstract}

\maketitle

\blfootnote{\textsuperscript{*}These authors contributed equally to this work.}

\section{Introduction}

The field of Quantum Computing has emerged with ever-increasing interest 
in both academia and industry. Quantum computers make use of the principles 
of quantum mechanics to solve computational problems that are difficult 
for classical computers \cite{ladd2010quantum, Chuang}. 
In recent years, quantum processors have grown substantially 
in the number of physical \textit{qubits} they support, enabling computations at increasingly larger scales \cite{kjaergaard2020superconducting,Saffman_2016}.
As a result, advances in quantum computing are creating new opportunities in fields such as
optimization \cite{quantum7010003}, machine learning \cite{10593221}, and the simulation of quantum phenomena in chemistry and materials science \cite{liuProspectsQuantumComputing2022}.

As in classical software, faults may arise during the design and implementation of quantum programs, creating a need for systematic techniques for assessing their correctness and reliability \cite{Fortunato2024, huangStatisticalAssertionsValidating2019}. However, testing quantum programs presents challenges not found in classical software. Quantum states cannot, in general, be directly inspected without measurement, and individual measurements provide only limited information about the underlying state while irreversibly disturbing it \cite{Chuang}. Furthermore, an unknown quantum state cannot be copied in general
~\cite{wootters1982single}, which further limits the ability to inspect and reuse intermediate program states during testing. An alternative is to compute the expected output through classical simulation by representing and simulating quantum states, which generally requires resources that grow exponentially with the number of qubits and are therefore infeasible in practical terms \cite{petabyte}. These characteristics make the oracle problem particularly relevant in quantum software testing: even when a quantum program can be executed, determining whether its output is correct may itself require substantial computational or experimental resources \cite{Fortunato2024, garcia2023quantum}.

Metamorphic testing (MT) offers a potential approach to alleviate this problem. Rather than requiring the expected output of an individual execution, MT verifies relations between multiple executions of the program \cite{chen2020metamorphictestingnewapproach}. Given a source test case, a metamorphic relation specifies how its input can be transformed into a follow-up input and how the corresponding outputs are expected to relate. This principle is already particularly appealing for quantum software, and previous work has already been developed in investigating metamorphic testing for certain quantum programs \cite{abreu_mt_oracle_2022} and quantum computing platforms \cite{paltenghi_morphq_2023} (see Section~\ref{sec:related}). However, given the previously stated effects of quantum mechanics, tailoring MT to a quantum setting introduces additional difficulties. For instance, given that quantum outputs cannot generally be inspected directly, a quantum metamorphic relation must also account for how the predicted relationship between outputs can be tested. Thus, simply replacing conventional test oracles with metamorphic relations does not by itself eliminate the difficulties associated with determining expected quantum behavior.   

In this work, we aim to bridge these differences by introducing a formal notion of quantum metamorphic testing (QMT) that explicitly accounts for the transformations required to measure quantum outputs. Rather than relying on generic measurement strategies that are independent of the metamorphic relations, our approach derives the measurement procedure for follow-up test cases from the corresponding source test case and metamorphic relation. This allows the testing process to exploit the structure of the metamorphic relation, potentially reducing the measurement effort required to assess the correctness of quantum programs if an efficient source measurement strategy has been derived.

We apply this notion of QMT to quantum programs aiming at the 
generation of stabilizer states. Stabilizer states form a 
particularly suitable family of quantum states, because they 
admit compact descriptions in terms of Pauli operators
\cite{gottesman1998heisenbergrepresentationquantumcomputers}. 
Consequently, their generation has been used to benchmark quantum 
computers~\cite{cao_generation_2023,cai2026generalizedmermininequalitiesbenchmarking}.
We focus in particular on graph states, a class of stabilizer states with a direct correspondence to mathematical graphs. The reason is that each stabilizer 
state is equivalent to a graph state up to simple local transformations which do not change entanglement, and therefore our approach adapts to stabilizer states in general. The graphical representation exposes structural transformations of the graph whose corresponding transformations of the quantum state are known \cite{hein2004multiparty}.
In particular, local complementation provides a graph transformation corresponding to local Clifford operations on the associated graph states \cite{vandennest_lc_2004}. 

We exploit these properties to derive metamorphic relations in which transformations of a graph, its associated quantum state, and the measurement settings for its verification
are used to test a program with the specification of preparing specific graph states. The considered metamorphic
relations preserve the main structure of the optimal verification measurements. This property reduces the experimental cost of performing these measurements on a quantum computer, as well as the cost of computing for every test case the suitable measurements.

In detail, we implement three composable metamorphic relations for graph-state preparation programs in Qiskit \cite{qiskit2024} and evaluate their efficiency and 
fault-detection effectiveness. Our evaluation shows a significant reduction in testing effort compared with testing independently generated random input graphs, while maintaining comparable fault detection effectiveness when the metamorphic relations are suitably combined. The approach also detects synthetic program faults introduced through circuit mutations and reveals a real bug in the Qiskit compiler.

The main contributions of this work are:
\begin{itemize}
\item A formalization of QMT that incorporates measurement procedures into metamorphic relations and supports their composition.
\item An application to graph-state preparation, with three composable relations that systematically derive follow-up inputs and their corresponding measurement circuits.
\item An empirical evaluation on Qiskit programs demonstrating reduced testing effort and the detection of both injected program faults and a real compiler bug.
\end{itemize}

\section{Background}

\subsection{Classical Metamorphic Testing}

Metamorphic Testing \cite{chen2020metamorphictestingnewapproach} is a testing methodology designed to alleviate the oracle problem, that is, situations in which determining the correct output of a program for arbitrary inputs is difficult or computationally expensive \cite{SeguraSurveyMetamorphicTesting2016}. This problem is particularly relevant in quantum computing, where obtaining the expected output may require costly simulations or analytical calculations. Instead of verifying individual outputs directly, MT checks whether multiple executions of the program satisfy expected relations (specifications) between their inputs and outputs \cite{LiuRevisitingEffectivenessMetamorphic2024, LiMetamorphicRelationGeneration2019}.

\begin{wrapfigure}[14]{r}{0.48\columnwidth}
    \vspace{-0.8\baselineskip}
    \centering

    \begin{tikzpicture}[
        >=stealth,
        thick,
        font=\huge,
        scale=0.6,
        transform shape
    ]

        \draw (0,0) circle (1.8cm);
        \draw (7,0) circle (1.8cm);

        \node at (-1.5,2.1) {$\mathcal{T}$};
        \node at (8.5,2.1) {$\mathcal{R}$};

        \coordinate (t)  at (-0.5,0.8);
        \coordinate (tk) at (0.5,-0.8);

        \coordinate (r)  at (6.5,0.8);
        \coordinate (rk) at (7.5,-0.8);

        \fill (t) circle (2pt);
        \node[left=5pt] at (t) {$t$};

        \fill (tk) circle (2pt);
        \node[left=5pt] at (tk) {$t'$};

        \fill (r) circle (2pt);
        \node[right=5pt] at (r) {$r$};

        \fill (rk) circle (2pt);
        \node[right=5pt] at (rk) {$r'$};

        \draw[->, bend left=15]
            ($(t)+(0.1,-0.1)$)
            to node[midway,right] {$I$}
            ($(tk)+(-0.1,0.1)$);

        \draw[->, bend left=15]
            ($(r)+(0.1,-0.1)$)
            to node[midway,right] {$O$}
            ($(rk)+(-0.1,0.1)$);

        \draw[->]
            ($(t)+(0.1,0)$)
            to node[midway,above] {$f$}
            ($(r)+(-0.1,0)$);

        \draw[->]
            ($(tk)+(0.1,0)$)
            to node[midway,above] {$f$}
            ($(rk)+(-0.1,0)$);

    \end{tikzpicture}

    \caption{\small
    A metamorphic transformation $I$ generates a follow-up test case
    $t'$ from a source test case $t$. The program $f$ maps the inputs
    from $\mathcal{T}$ to outputs in $\mathcal{R}$. The output mapping
    $O$ determines the expected relation between the source output
    $r$ and follow-up output $r'$.}
    \label{fig:metamorphic-relation}

   \end{wrapfigure}
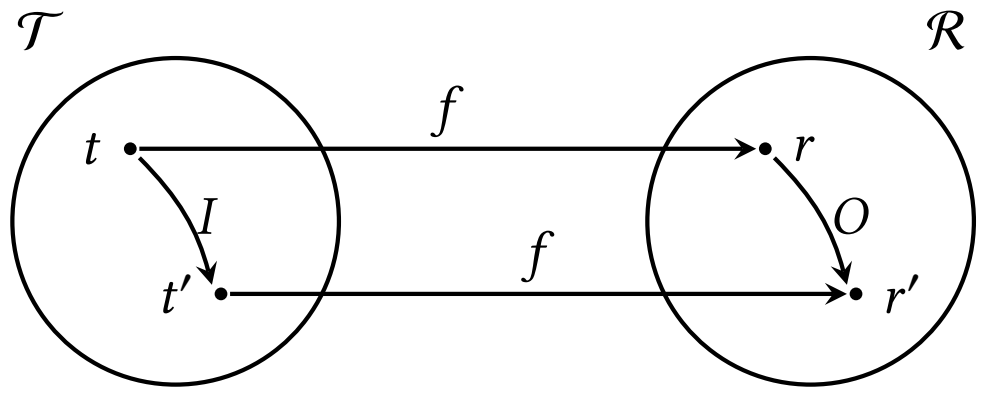

As a simple example, consider a program intended to compute the sine function for an input \(x\). We now consider the identity
\[
\sin(x+ (2k+1) \cdot \pi) = -\sin(x),
\]
that holds for all \(x\) with \(k \in \mathbb{Z}\). A correct implementation of the sine function should return outputs for inputs (\(x\)) and (\(x+ (2k+1) \cdot \pi\)) that differ only by a sign change. This expected relationship between inputs and outputs naturally defines a metamorphic relation (MR) \cite{chen2020metamorphictestingnewapproach}. Thus, let \( f : \mathcal{T} \rightarrow \mathcal{R} \) be a program under test, where \(\mathcal{T}\) denotes the input domain and \(\mathcal{R}\) the output domain. Following the standard formulation of metamorphic testing in the literature, we formalize metamorphic relations in Definition \ref{def:MR}; also depicted in Figure \ref{fig:metamorphic-relation}.

\begin{definition}[Metamorphic relation]
  \label{def:MR}
  Let $f \colon \mathcal{T} \rightarrow \mathcal{R}$ be a program and
  $T \subseteq \mathcal{T}$ a set of source inputs. A \emph{metamorphic
  relation} $\mathrm{MR} = (I, O)$ consists of
  \begin{enumerate}
    \item an \emph{input mapping} $I \colon T \rightarrow \mathcal{T}$,
          which generates the follow-up input $t' = I(t)$, and
    \item an \emph{output mapping} $O \colon \mathcal R \rightarrow \mathcal{R}$,
          which maps the source output $f(t)$ to the expected follow-up
          output,
  \end{enumerate}
  such that every correct implementation of $f$ satisfies
  \begin{equation}
    f\bigl(I(t)\bigr) = O\bigl(f(t)\bigr)
    \qquad \text{for all } t \in T .
    \label{eq:MR}
  \end{equation}
\end{definition}

In practice, a program typically satisfies not just a single metamorphic relation but an entire family $\{\mathrm{MR}_k\}_k$, where each $\mathrm{MR}_k = (I_k, O_k)$ with set of source inputs $T_k$ is an instance of \cref{def:MR}.

Returning to the sine example, each integer $k$ defines a metamorphic relation with $I_k(x)=x+(2k+1)\pi$ and $O_k(r)=-r$. Given a source input $x$, we execute the program to obtain $r=f(x)$, generate the follow-up input $x'=x+(2k+1)\pi$, and check whether
\vspace{0.5em}
\par
\noindent\hfill \(f(x')=-f(x).\)\hfill\mbox{}
\par
\vspace{0.5em}
A violation indicates a fault, whereas satisfying the relation does not prove correctness. For instance, a program that always returns zero satisfies this relation for every input, which motivates testing multiple source inputs and different metamorphic relations.

Moreover, several metamorphic relations can be applied in sequence to generate a single follow-up test case, provided that each generated input is a valid source input for the next relation. This process is known as \emph{metamorphic relation composition}~\cite{LiuRevisitingEffectivenessMetamorphic2024}. The input mappings are applied successively to the source input, while the corresponding output mappings are applied in the same order to the source output to derive the expected final output. Thus, composing $(I_1,O_1)$ and $(I_2,O_2)$ produces the follow-up input $I_2(I_1(x))$ and expected output $O_2(O_1(f(x)))$. 

For example, applying the sine relation first with $k=0$ and then with $k=1$ transforms $x$ into $x+4\pi$. Since each relation flips the sign of the expected output, the two sign changes cancel, yielding 
\par\vspace{0.5em}
\noindent\hfill$f(x+4\pi)=f(x).$\hfill\refstepcounter{equation}(\theequation)\label{eq:sine-composition}
\par\vspace{0.5em}

However, the challenging aspect of Metamorphic Testing lies in identifying relevant metamorphic relations from knowledge of the program under test. These relations should capture properties that the program must satisfy according to its specification.

\WFclear
The main advantage of this method lies in the ability to derive new test cases from a source test case, rather than constructing them from scratch. Generating new valid test cases can be an expensive and labor-intensive task, and test case generation has a strong impact on the effectiveness and efficiency of software testing \cite{ANAND20131978}. However, given appropriate metamorphic relations, valid follow-up test cases can be systematically derived from existing source test cases and thus ease the testing effort. 

\subsection{Fundamental Concepts of Quantum Computing}
\label{subsec:fundamentals_QC}
In the circuit model of quantum computing (\cref{fig:bell:circuit}), the program state is encoded in quantum bits, or \emph{qubits}; operations on this state are represented by unitary operators, or gates; and the final state is read out through measurements typically at the end of the computation \cite{Chuang}. Qubits are initialized all at the same state, and then operated by the gates and measurements whose probabilistic outcomes are stored in classical registers. In order to perform quantum computation in practice, transpilation is needed to translate a desired circuit into available hardware with a restricted set of quantum gates. One needs to take into account the topology of interaction between qubits, which can affect implementation and demands tailored optimization for each protocol. For instance, it may be  that not all qubits can be measured or operated simultaneously, and implementations require engineering optimal interactions. Widely used frameworks such as Qiskit \cite{qiskit2024}, PennyLane \cite{bergholm2022pennylane}, and Cirq \cite{Cirq_2026} implement this model, allowing developers to compose circuits from these elements and execute them on classical simulators or quantum hardware (\cref{fig:bell:qiskit}). In this section, we introduce these fundamental building blocks of quantum programs.

In classical computation, the basic unit of information, the \emph{bit}, can take one of two possible values, $0$ or $1$. A qubit, by contrast, can occupy any coherent superposition of the two computational basis states denoted by $\ket{0}$ and $\ket1$:
\[ \ket{\psi} = \alpha\ket{0}+\beta\ket1,\]
where the complex probability amplitudes $\alpha,\beta\in\mathbb{C}$ satisfy the normalization condition $|\alpha|^2+|\beta|^2=1$. For a system of $N$ qubits, the computational basis consists of tensor products of the single-qubit basis states,
\[ \ket x := \ket{x_1}\otimes \ket{x_2}\otimes \cdots \otimes \ket{x_N}, \]
where $x=(x_1,\ldots,x_N)\in\{0,1\}^N$ is a bit string of length $N$. Equivalently, $x$ may be identified with the integer represented by this binary string. A general $N$-qubit state can therefore be written as
\[
\ket \psi = \sum_{x = 0}^{2^N-1} c_x \ket{x},
\]
where the amplitudes $c_x\in\mathbb{C}$ satisfy $\sum_{x=0}^{2^N-1}|c_x|^2=1$. A state is called \emph{entangled} if it cannot be written as a tensor product of single-qubit states, i.e., if there are no states $\ket{\psi_1}, \ldots, \ket{\psi_N}$ such that $\ket{\psi} = \ket{\psi_1} \otimes \cdots \otimes \ket{\psi_N}$. A standard example is the Bell state
of two qubits, $ \ket{\Phi^+}=(\ket{00} + \ket{11})/\sqrt{2}$. By convention, the system is typically initialized in the state $\ket{00\ldots 0}$, meaning that every qubit is in the state $\ket{0}$ (see \cref{fig:bell}).

\begin{figure}[t]
		\centering
		\begin{subfigure}[c]{0.47\textwidth}
			\centering
			\begin{quantikz}[row sep=0.3cm, column sep=0.35cm]
				\lstick{$q_1:\ket{0}$} & \gate{H} & \ctrl{1} & \meter{} & \cw \rstick{$c_1$} \\
				\lstick{$q_2:\ket{0}$} & \qw      & \targ{}  & \meter{} & \cw \rstick{$c_2$}
			\end{quantikz}
			\caption{Quantum circuit.}
			\label{fig:bell:circuit}
		\end{subfigure}
		\hfill
		\begin{subfigure}[c]{0.49\textwidth}
			\centering
			\begin{lstlisting}[style=qiskit]
from qiskit import QuantumCircuit
				
# q_1 -> 0, q_2 -> 1; 2 classical bits
qc = QuantumCircuit(2, 2)
# state: |00>   (written as |q_1 q_2>)

qc.h(0)      # H on q_1
# state: (|00>+|10>)/sqrt(2)

qc.cx(0, 1)  # CNOT: c=q_1, t=q_2
# state: (|00>+|11>)/sqrt(2)

# measure q_1 -> c_1 and q_2 -> c_2
qc.measure([0, 1], [0, 1])
\end{lstlisting}
			\caption{Qiskit implementation \cite{qiskit2024}.}
			\label{fig:bell:qiskit}
		\end{subfigure}
		\caption{Preparation and measurement of the maximally entangled Bell state
			$\ket{\Phi^+}=(\ket{00}+\ket{11})/\sqrt2$. (\subref{fig:bell:circuit}) circuit diagram: one initializes qubits $q_1$ and $q_2$, applies a Hadamard gate $H$ on $q_1$, and a CNOT gate with control $q_1$ and target $q_2$, and then measures the prepared state in the $Z$ basis at each register. The measurement
            outcomes are encoded in two classical bits $c_1$ and $c_2$.
			(\subref{fig:bell:qiskit}) the corresponding Qiskit code to execute this circuit. Qiskit indexes qubits from 0, so $q_i$ corresponds to index $i-1$ in the code.}
		\label{fig:bell}
\end{figure}

A \emph{quantum gate} is a unitary operator $U$, i.e., one satisfying $UU^\dagger = U^\dagger U = \one$, where $U^\dagger$ denotes the conjugate transpose of $U$ and $\one$ is the identity operator. Unitarity implies that quantum gates are reversible and preserve the norm of the states on which they act. Identifying the computational basis states $\ket{x}$ with the standard basis vectors, we represent gates by unitary matrices. Important examples of single-qubit gates are the Hadamard gate $H$, which maps computational basis states to equal-weight superpositions, and the phase gates $S$ and $T$, which introduce relative phases between $\ket{0}$ and $\ket{1}$:
\begin{equation}
H = \frac{1}{\sqrt{2}}
\begin{pmatrix}
1 & 1 \\
1 & -1
\end{pmatrix},
\qquad
S =
\begin{pmatrix}
1 & 0 \\
0 & i
\end{pmatrix},
\qquad
T =
\begin{pmatrix}
1 & 0 \\
0 & e^{i\pi/4}
\end{pmatrix}.
\end{equation}
The two-qubit gate $\operatorname{CNOT}_{c,t}$, with control qubit $c$ and
target qubit $t$, acts on the computational basis states analogously to the
classical controlled-NOT gate:
\begin{equation}
\operatorname{CNOT}_{c,t}\ket{x}_c\ket{y}_t = \ket{x}_c\ket{x \oplus y}_t,
\end{equation}
where $\oplus$ denotes addition modulo~2. That is, the target qubit is flipped
if and only if the control qubit is in the state $\ket{1}$. Unlike single-qubit gates, $\operatorname{CNOT}$ is an \emph{entangling} gate: it can map product states to entangled states. For example, the circuit in \cref{fig:bell}, which applies a Hadamard gate to the control qubit followed by a $\operatorname{CNOT}$, produces the Bell state
\begin{equation}
  \operatorname{CNOT}_{1,2} (H \otimes \one) \ket{00}
  = \operatorname{CNOT}_{1,2} \frac{\ket{00} + \ket{10}}{\sqrt{2}}
  = \frac{\ket{00} + \ket{11}}{\sqrt{2}}.
\end{equation}
The gate set $\{H, S, \operatorname{CNOT}\}$ generates the Clifford group
and suffices to generate stabilizer states, but is not universal. The set
$\{H, T, \operatorname{CNOT}\}$ is, however, universal:
Any unitary operation on
$N$ qubits can be approximated to arbitrary precision by a finite circuit
composed of these gates \cite{Chuang}. A unitary $U$ is \emph{local} if it decomposes as a tensor product of single-qubit unitaries $U= U_1\otimes U_2\otimes \cdots \otimes U_N$.

\medskip

Unlike a classical bit, a quantum state cannot simply be read out. Instead, information about it must be extracted by measurements, which in general disturbs the state \cite{Chuang}. For a single qubit, the most important single-qubit measurements are those associated with the Pauli operators
\begin{equation}
  X = \begin{pmatrix} 0 & 1 \\ 1 & 0 \end{pmatrix}, \qquad
  Y = \begin{pmatrix} 0 & -i \\ i & 0 \end{pmatrix}, \qquad
  Z = \begin{pmatrix} 1 & 0 \\ 0 & -1 \end{pmatrix}.
  \label{eq:pauli}
\end{equation}
Together with the identity $\one$, they form a basis of all 
single-qubit operators. Each Pauli operator is Hermitian and 
unitary, squares to the identity, and has eigenvalues $\pm 1$, 
so it can serve both as a gate and as an observable. Measuring 
a Pauli operator in the state $\ket{\psi}$ results in  
one of the eigenvalues $+1$ or $-1$ as potential outcomes;
each outcome occurs with the probability 
$p_\pm = |\braket{\psi}{\phi_\pm}|^2$, where $\ket{\phi_\pm}$
are the corresponding eigenvectors. Concretely, the eigenstates of 
$Z$ are the computational basis states, $Z\ket{0} = \ket{0}$ 
and $Z\ket{1} = -\ket{1}$, so a computational-basis measurement 
is a $Z$ measurement. The eigenstates of $X$ are 
$\ket{\pm} = (\ket{0} \pm \ket{1})/\sqrt{2}$, and those 
of $Y$ are $\ket{\pm i} = (\ket{0} \pm i\ket{1})/\sqrt{2}$, 
each with eigenvalue $\pm 1$. For a multi-qubit state 
$\ket{\psi} = \sum_x c_x \ket{x}$, a measurement in the 
computational basis corresponds to measuring $Z$ on all qubits, 
and yields outcome $x$ with probability $p(x) = |c_x|^2$, 
and the post-measurement state is $\ket{x}$.

Measurements in other bases are typically realized by applying a basis-change unitary $U$ before a measurement in the computational basis. Outcome $x$ then occurs with probability
\begin{equation}
  p(x) = |\bra{x} U \ket{\psi}|^2 = |\braket{b_x}{\psi}|^2,
  \qquad \ket{b_x} = U^\dagger \ket{x},
\end{equation}
which corresponds to a measurement in the basis $\{\ket{b_x}\}$. Measuring $X$ requires $U = H$, since $H\ket{+} = \ket{0}$ and $H\ket{-} = \ket{1}$. Measuring $Y$ requires $U = H S^\dagger$, i.e., $S^\dagger$ followed by $H$, since $S^\dagger\ket{\pm i} = \ket{\pm}$. In all three cases, outcome $0$ corresponds to the eigenvalue $+1$ and outcome $1$ to the eigenvalue $-1$ of the respective Pauli operator. Measuring in all $3^N$ combinations of the $X$, $Y$,
and $Z$ bases on the individual qubits, with sufficiently many shots each,
yields a complete characterization of the state. However, the exponential
number of measurement settings, combined with the shots required for each,
quickly renders this approach infeasible as the number of qubits grows.
Efficient tests must therefore rely on well-chosen
measurements that exploit prior knowledge of the expected state. This
motivates adapting established classical testing techniques, such as
metamorphic testing, to the quantum setting.

\section{Quantum Metamorphic Testing}
In this section, we extend metamorphic testing to quantum software. 
Here, we consider the situation where a quantum program is designed to 
result in a specific quantum state.
This may be the final output of a program or an intermediate state at 
some point in the circuit. 
Even in cases where an efficient classical simulation is possible
~\cite{gottesman1998heisenbergrepresentationquantumcomputers}, the practical testing
of the resulting state remains challenging for large systems. In addition, hardware 
execution exposes faults that are invisible in simulation, 
such as errors introduced during transpilation, qubit mapping, or calibration.
In any quantum hardware, state preparation can be verified only through measurements. So, the goal is to exploit (partial) prior knowledge of 
the target state to construct tests that efficiently certify either 
the state itself or selected properties of it. Here, the notion of metamorphic relations can become relevant: If the input of the quantum
program is changed in a specific way, the measurement strategies needed
for the state testing can sometimes be adapted straightforwardly, allowing 
to generate more test-cases in an efficient manner.

If a given quantum state should be characterized, it is preferable if it 
can be tested by local measurements with deterministic outcomes. These are 
considerably easier to handle than a given expectation value $\expval{M}$ 
of a non-deterministic measurement $M$. In fact, in the deterministic case, 
only a single outcome is permitted, so observing any other outcome immediately 
reveals a fault, and the probability of missing a fault decreases exponentially 
with the number of shots. By contrast, verifying $\expval{M}$ requires estimating 
it from the sample mean of many measurement outcomes, which for a bounded observable 
requires on the order of  $1/\varepsilon^2$ shots to reach precision $\varepsilon$, 
and any finite sample leaves residual statistical uncertainty~\cite{hoeffding_1963}.
Stabilizer states are a natural scenario where measurement strategies are simple. 
By definition, stabilizer states are multi-qubit states which are eigenstates of 
certain tensor products of Pauli matrices, resulting directly in the desired local
measurements with deterministic outputs. This enables significantly more efficient 
software testing.

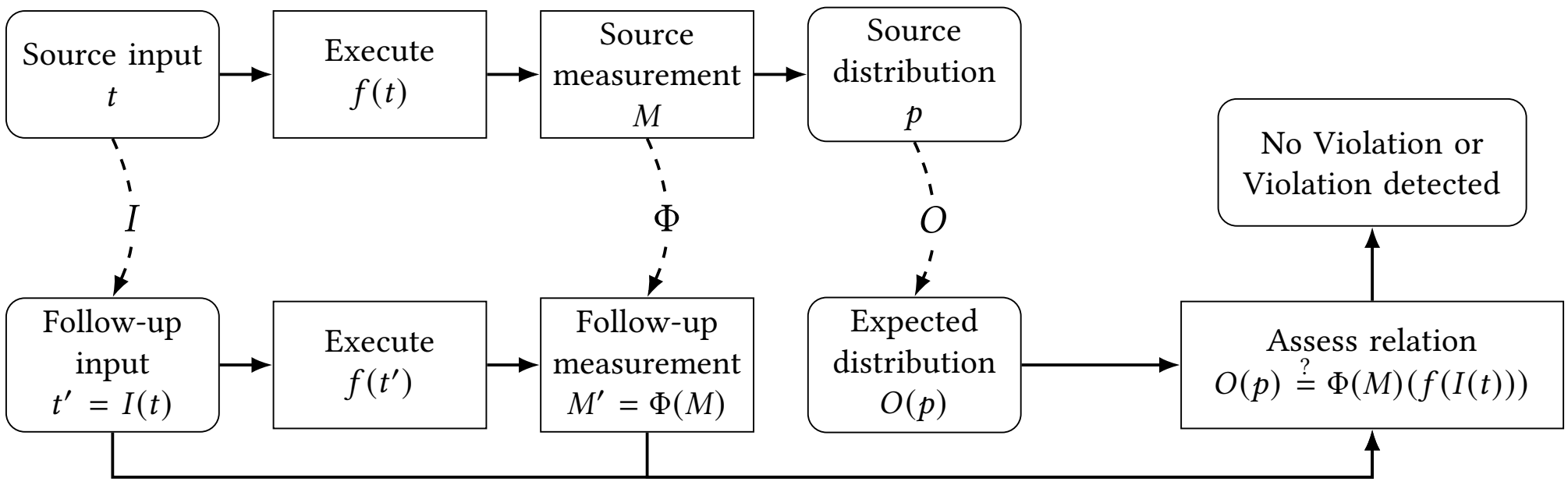
\begin{figure*}[t]
\centering
\begin{adjustbox}{max width=0.95\linewidth}
\begin{tikzpicture}[
    font=\small,
    box/.style={
        draw,
        text width=18mm,
        minimum height=12mm,
        align=center,
        inner sep=3pt
    },
    data/.style={
        box,
        rounded corners=5pt
    },
    process/.style={
        box
    },
    flow/.style={
        -{Latex[length=2mm]},
        thick
    },
    mapping/.style={
        -{Latex[length=2mm]},
        dashed,
        thick
    },
    maplabel/.style={
        font=\normalsize,
        fill=white,
        inner sep=2pt
    }
]

\node[data] (input)
    {Source input\\$t$};

\node[process, right=5mm of input] (program)
    {Execute\\$f(t)$};

\node[process, right=5mm of program] (measure)
    {Source\\measurement\\$M$};

\node[data, right=5mm of measure] (output)
    {Source\\distribution\\$p$};

\node[data, below=15mm of input] (input2)
    {\mbox{Follow-up}\\input\\$t'=I(t)$};

\node[process, right=5mm of input2] (program2)
    {Execute\\$f(t')$};

\node[process, right=5mm of program2] (measure2)
    {Follow-up\\measurement\\$M'=\Phi(M)$};

\node[data, right=5mm of measure2] (output2)
    {Expected\\distribution\\$O(p)$};

\draw[flow] (input.east) -- (program.west);
\draw[flow] (program.east) -- (measure.west);
\draw[flow] (measure.east) -- (output.west);

\draw[flow] (input2.east) -- (program2.west);
\draw[flow] (program2.east) -- (measure2.west);

\draw[mapping] (input.south)
    to[bend left=25]
    node[maplabel, midway] {\large $I$}
    (input2.north);

\draw[mapping] (measure.south)
    to[bend left=25]
    node[maplabel, midway] {\large $\Phi$}
    (measure2.north);

\draw[mapping] (output.south)
    to[bend left=25]
    node[maplabel, midway] {\large $O$}
    (output2.north);

\coordinate (laneMidpoint) at
    ($(output.center)!0.5!(output2.center)$);

\node[process, text width=34mm, right=15mm of output2]
    (compare)
    {Assess relation\\$O(p) \stackrel{?}{=} \Phi(M)(f(I(t)))$};

\coordinate (upperEntry) at
    ([yshift=3mm]compare.west);

\coordinate (lowerEntry) at
    ([yshift=-3mm]compare.west);

\coordinate (routeMidpoint) at
    ($(output.east)!0.5!(compare.west)$);

\draw[flow]
    (input2.south)
    -- ++(0pt,-12pt)
    -- (compare.south |- {$(input2.south)+(0pt,-12pt)$})
    -- (compare.south);

    \draw[flow]
    (measure2.south)
    -- ++(0pt,-12pt)
    -- (compare.south |- {$(input2.south)+(0pt,-12pt)$})
    -- (compare.south);
    
\draw[flow]
    (output2.east)
    -- (compare.west);

\node[data, text width=27mm, above=7mm of compare] (verdict)
    {No Violation or \\
     Violation detected};

\draw[flow] (compare.north) -- (verdict.south);

\end{tikzpicture}
\end{adjustbox}

\caption{General quantum metamorphic testing pipeline.
Solid arrows denote dataflow, while dashed arrows denote
metamorphic mappings. The mappings $I$, $\Phi$, and $O$ derive
the follow-up input, measurement strategy, and expected output
distribution, respectively. The expected follow-up distribution
$O(p)$ is then compared with the observed distribution of the follow-up input $\Phi(M)(f(I(t)))$.}
\label{fig:qmt-pipeline}
\end{figure*}

More generally, the classical definition of MT has to be extended for quantum outputs to also include transformations of the measurement strategies, i.e., of the mappings from the quantum state to probability distributions. In the following, we restrict ourselves to pure states and model a quantum program as a map $f \colon \mathcal{T} \rightarrow \mathcal{H}$ from classical inputs to normalized states $\ket{\psi} \in \mathcal{H}$ of the $N$-qubit Hilbert space $\mathcal{H}$. As described in \cref{subsec:fundamentals_QC}, a measurement in an arbitrary basis is implemented by a sequence of gates, described by a unitary $U$, followed by a measurement in the computational basis which yields a probability distribution over the possible outcomes. The raw outcomes may additionally be post-processed classically, e.g., by computing the parity of a subset of bits to obtain the eigenvalue of a Pauli string. To cover all such cases, we describe a \emph{measurement strategy} in general as a map $M \colon \mathcal{H} \rightarrow \mathcal{P}$, where $\mathcal{P}$ denotes the set of probability distributions over a finite set of outcomes $\Omega$ (see \cref{fig:qmt-pipeline}).

\begin{definition}[Quantum metamorphic relation]
  \label{def:QMR}
  Let $f \colon \mathcal{T} \rightarrow \mathcal{H}$ be a quantum program,
  $T \subseteq \mathcal{T}$ a set of source inputs, and $\mathcal{M}$ a set
  of measurement strategies $M \colon \mathcal{H} \rightarrow \mathcal{P}$.
  A \emph{quantum metamorphic relation} $\mathrm{QMR} = (I, \Phi, O)$
  consists of
  \begin{enumerate}
    \item an \emph{input mapping} $I \colon T \rightarrow \mathcal{T}$,
          which generates the follow-up input $t' = I(t)$,
    \item a \emph{measurement mapping}
          $\Phi \colon \mathcal{M} \rightarrow \mathcal{M}'$, which assigns
          to every source measurement strategy $M$ a follow-up measurement
          strategy $M' = \Phi(M)$, where $\mathcal{M}'$ is a set of
          measurement strategies, and
    \item an \emph{output mapping} $O \colon \mathcal{P} \rightarrow \mathcal{P}$,
          which maps the source outcome distribution to the expected
          follow-up outcome distribution,
  \end{enumerate}
  such that every correct implementation of $f$ satisfies
  \begin{equation}
    \Phi(M)\bigl(f(I(t))\bigr) = M'\bigl(f(I(t))\bigr) = O\bigl(M(f(t))\bigr)
    \qquad \text{for all } t \in T,\ M \in \mathcal{M} .
    \label{eq:QMR}
  \end{equation}
\end{definition}

\Cref{def:QMR} contains the classical case of \cref{def:MR} as a special
case. Suppose that the program outputs classical data, i.e.,
$f(t) = \ket{f_{\mathrm{c}}(t)}$ is a computational basis state for a
classical program $f_{\mathrm{c}} \colon \mathcal{T} \rightarrow \{0,1\}^N$.
Choosing $\mathcal{M} = \{M\}$ with $M$ the plain computational-basis
measurement, $\Phi(M) = M$, and the output mapping induced by a classical
output mapping $O_{\mathrm{c}}$, i.e., $O(\delta_r) = \delta_{O_{\mathrm{c}}(r)}$,
where $\delta_r$ denotes the distribution concentrated on outcome $r$, we
obtain $M(f(t)) = \delta_{f_{\mathrm{c}}(t)}$. \Cref{eq:QMR} then reads
$\delta_{f_{\mathrm{c}}(I(t))} = \delta_{O_{\mathrm{c}}(f_{\mathrm{c}}(t))}$,
which holds if and only if
$f_{\mathrm{c}}(I(t)) = O_{\mathrm{c}}(f_{\mathrm{c}}(t))$, i.e., the
classical relation of \cref{eq:MR}.

Composition carries over to quantum metamorphic relations, with the
additional requirement that the measurement mappings must be composable as
well: the follow-up measurement strategies of the first relation have to be
valid source strategies of the second.

In the following, we demonstrate quantum metamorphic testing for the verification of graph states, an important subclass of stabilizer states. Graph states are particularly well suited for this purpose: the graph transformations introduced in the next subsection provide metamorphic relations whose follow-up states are known exactly, and there is a canonical way of constructing Pauli measurements which yield measurements with deterministic outcomes. Together, these properties allow us to construct efficient tests suited for the execution on quantum hardware.

\subsection{Quantum Metamorphic Testing of Stabilizer States}
\label{sec:approach}

Our main goal is to formulate a QMT framework to test the preparation of
stabilizer states, which are the main resources for protocols like measurement-based quantum computing~\cite{briegel2001persistent,raussendorf2001oneway} or quantum error correction~\cite{schlingemann2001quantum}. For general quantum programs, computing the expected output is itself costly, since classical simulation typically scales exponentially with the number of qubits. This obstacle is partly avoided for circuits consisting only of stabilizer-state preparation, Clifford gates, and Pauli measurements, as these can be simulated efficiently on classical hardware~\cite{gottesman1998heisenbergrepresentationquantumcomputers}. However, simulation alone does not tell us which measurements to perform on the device, and finding suitable measurements efficiently for each test case remains a central challenge for practical verification~\cite{eisert2020certification}. In particular, practical hardware can only perform few-qubit measurements, and it is difficult to find what these measurements are for a given stabilizer state. To overcome this limitation when certifying stabilizer states we will use that, up to relabeling of the Pauli matrices, stabilizer states can be described by graph states. Therefore, the program under test receives a graph $G$ as input and is
supposed to produce the corresponding graph state $\ket{G}$. The structure of graph states provides natural metamorphic relations:
qubit permutations, as well as certain local Clifford
unitaries and $\mathrm{CNOT}$
gates (possibly followed by a $Z$
gate) map graph states to graph states \cite{vandennest_lc_2004,Gachechiladze_2017UnitaryGraph}. Building upon these relations, we develop an efficient framework to test such states from the corresponding few-qubit measurements, without computing them for every test case. This framework is presented below.

\bigskip

Graph states are described by mathematical graphs $G = (V, E)$, whose vertices $i \in V$
represent qubits and whose edges $\{i, j\} \in E$ represent controlled-$Z$ (CZ)
gates between the corresponding qubits \cite{hein2004multiparty}. Starting from all qubits in the
state $\ket{+}$, the graph state $\ket{G}$ is obtained by applying a
$\mathrm{CZ}$ gate for every edge, where
$\mathrm{CZ}\ket{x}\ket{y} = (-1)^{xy}\ket{x}\ket{y}$.
Graph states are associated with a set of measurements with deterministic outcomes, these form the elements of their so-called stabilizer. This group is constructed from the graph as follows: For each vertex $i$ one considers the operator $K_i = X_i \otimes_{j \in \mathcal{N}(i)} Z_j$, where $\mathcal{N}(i)$ denotes
the neighborhood of $i$ and the identity operator on the remaining qubits left implicit. For example, for the graph 
in Fig. 4(a) one has $K_1 = X_1 Z_2$, 
$K_2 = Z_1 X_2 Z_3$, and 
$K_3 = Z_2 X_3$.
Then, $K_i \ket{G}=\ket{G}$ or equivalently $\bra{G} K_i \ket{G} = 1$. The $K_i$ and their products form an abelian group, the stabilizer.
Each element $s$ is a Pauli string up to a sign, i.e.,
$s = \pm P_1 \otimes P_2 \otimes \cdots \otimes P_N$ with
$P_j \in \{\one, X, Y, Z\}$. Measuring $s$ requires single-qubit Pauli
measurements on the qubits in its support, i.e., on all qubits $j$ with
$P_j \neq \one$. The number of tensor factors different from the identity $\one$ is the Pauli weight of $s$.
Here, we restrict ourselves to stabilizer elements of weight at most $w_{\max}$
for two reasons. First, photon loss on photonic platforms makes
measurements involving many qubits increasingly unlikely to succeed.
Second, low-weight measurements are more robust to readout errors. The
drawback of this restriction is that finding low-weight stabilizer elements
is computationally hard. The minimum weight of a non-trivial stabilizer 
element of $\ket{G}$ equals one plus the \emph{local minimum degree} of $G$,
i.e., the smallest vertex degree attainable through a sequence of local 
complementations~\cite{hoyer_resources_2006}, and the corresponding decision
problem is NP-complete~\cite{javelle_minimum_2012}. We alleviate this
problem with the metamorphic relations presented in the following: instead
of searching for low-weight stabilizer elements of each follow-up state,
we derive its deterministic measurements from known measurements of the
source state via the metamorphic relations \emph{PERM}, \emph{LC}, and \emph{CNOT} presented next.

\paragraph{PERM} The simplest quantum metamorphic relation is based on permuting the qubits. Relabeling the vertices of the input graph should produce the same
graph state, with the roles of the qubits exchanged accordingly. A correct
program therefore yields a follow-up state that differs from the source
state only in the order of its qubits. The expected measurements of the
follow-up state are obtained by relabeling those of the source state in the
same way, so they remain deterministic, keep their weight, and have the
same expected outcomes. Although simple, this relation can reveal faults
that depend on the specific assignment of qubits, for instance errors in
how the program maps graph vertices to qubits, or faults affecting
particular qubits or connections on the hardware.

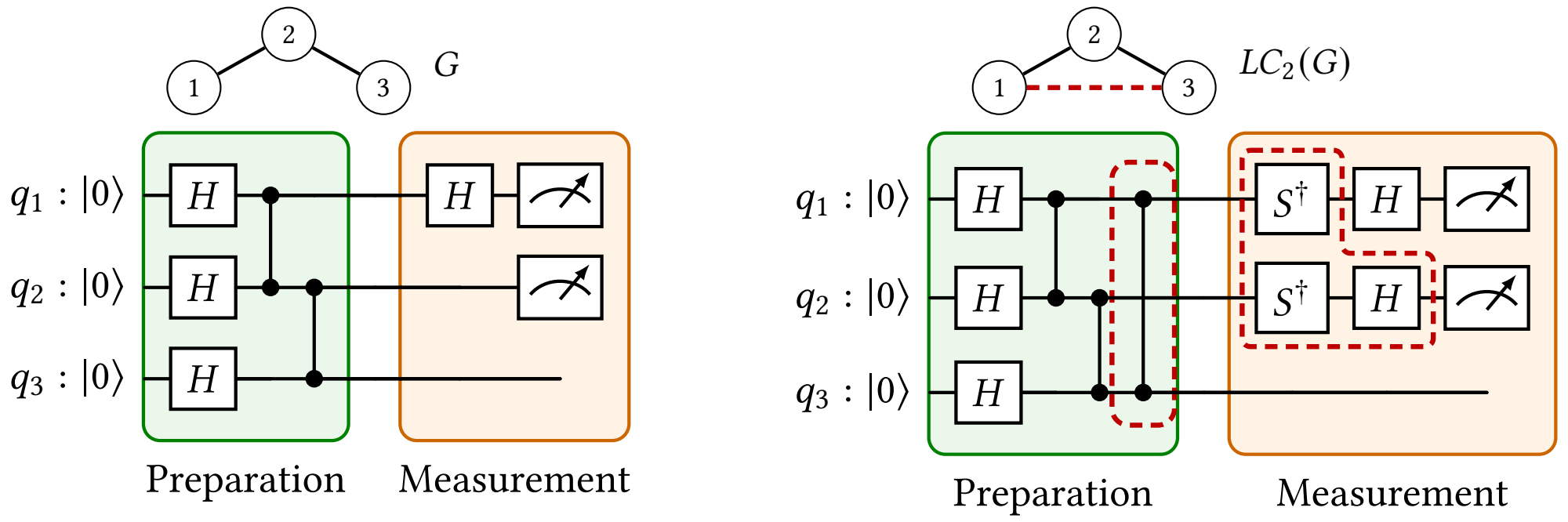
\begin{figure*}[t]
    \centering

    \begin{subfigure}[t]{0.485\textwidth}
        \centering
        \begin{tikzpicture}[
            vertex/.style={circle, draw, minimum size=4.5mm, inner sep=0pt,
                           font=\scriptsize, fill=white}
        ]
            \node[vertex] (q1) at (0.8,0)      {1};
            \node[vertex] (q2) at (1.6,0.45) {2};
            \node[vertex] (q3) at (2.4,0)    {3};
            \draw[thick] (q1) -- (q2);
            \draw[thick] (q2) -- (q3);
            \node[anchor=west] at (2.7,0.2) {\small $G$};
        \end{tikzpicture}

        \vspace{1mm}

        \begin{quantikz}[row sep=0.25cm, column sep=0.22cm]
            \lstick{$q_1:\ket{0}$}
                & \gate{H} \gategroup[
                    wires=3, steps=3,
                    style={rounded corners, draw=green!50!black, fill=prepbox, inner xsep=3pt},
                    background,
                    label style={label position=below, anchor=north, yshift=-0.2cm}
                  ]{\strut Preparation}
                & \ctrl{1}
                & \qw
                & \qw & \qw & \qw
                & \gate{H} \gategroup[
                    wires=3, steps=2,
                    style={rounded corners, draw=orange!80!black, fill=measbox, inner xsep=3pt},
                    background,
                    label style={label position=below, anchor=north, yshift=-0.2cm}
                  ]{\strut Measurement}
                & \meter{} \\
            \lstick{$q_2:\ket{0}$}
                & \gate{H} & \control{} & \ctrl{1}
                & \qw & \qw & \qw
                & \qw & \meter{} \\
            \lstick{$q_3:\ket{0}$}
                & \gate{H} & \qw & \control{}
                & \qw & \qw & \qw
                & \qw & \qw
        \end{quantikz}

        \caption{Source test case for the graph $G:1$--$2$--$3$, with the
        measurement circuit for the stabilizer element $K_1 = X_1 Z_2$.}
        \label{fig:lc-example-source}
    \end{subfigure}
    \hfill
    \begin{subfigure}[t]{0.485\textwidth}
        \centering
        \begin{tikzpicture}[
            vertex/.style={circle, draw, minimum size=4.5mm, inner sep=0pt,
                           font=\scriptsize, fill=white}
        ]
            \node[vertex] (q1) at (0.8,0)      {1};
            \node[vertex] (q2) at (1.6,0.45) {2};
            \node[vertex] (q3) at (2.4,0)    {3};
            \draw[thick] (q1) -- (q2);
            \draw[thick] (q2) -- (q3);
            \draw[red!75!black, very thick, densely dashed] (q1) -- (q3);
            \node[anchor=west] at (2.7,0.2) {\small $LC_2(G)$};
        \end{tikzpicture}

        \vspace{1mm}

        \begin{quantikz}[row sep=0.25cm, column sep=0.22cm,
            execute at end picture={
                \draw[red!75!black, very thick, densely dashed, rounded corners=2pt]
                    ($(sdtop.north west)+(-3pt,3pt)$)
                    -- ($(sdtop.north east)+(3pt,3pt)$)
                    -- ($(sdtop.east |- hbottom.north)+(3pt,3pt)$)
                    -- ($(hbottom.north east)+(3pt,3pt)$)
                    -- ($(hbottom.south east)+(3pt,-4pt)$)
                    -- ($(sdbottom.south west)+(-3pt,-3pt)$)
                    -- cycle;
            }]
            \lstick{$q_1:\ket{0}$}
                & \gate{H} \gategroup[
                    wires=3, steps=4,
                    style={rounded corners, draw=green!50!black, fill=prepbox, inner xsep=3pt},
                    background,
                    label style={label position=below, anchor=north, yshift=-0.2cm}
                  ]{\strut Preparation}
                & \ctrl{1}
                & \qw
                & \ctrl{2}
  \gategroup[
      wires=3,
      steps=1,
      style={
          draw=red!75!black,
          very thick,
          densely dashed,
          rounded corners,
          inner xsep=-10pt,
          inner ysep=-3pt
      },
      background
  ]{}
                & \qw & \qw & \qw
                & \gate[style={name=sdtop}]{S^\dagger} \gategroup[
                    wires=3, steps=3,
                    style={rounded corners, draw=orange!80!black, fill=measbox, inner xsep=3pt},
                    background,
                    label style={label position=below, anchor=north, yshift=-0.2cm}
                  ]{\strut Measurement}
                & \gate{H}
                & \meter{} \\
            \lstick{$q_2:\ket{0}$}
                & \gate{H} & \control{} & \ctrl{1} & \qw
                & \qw & \qw & \qw
                & \gate[style={name=sdbottom}]{S^\dagger}
                & \gate[style={name=hbottom}]{H}
                & \meter{} \\
            \lstick{$q_3:\ket{0}$}
                & \gate{H} & \qw & \control{} & \control{}
                & \qw & \qw & \qw
                & \qw & \qw & \qw
        \end{quantikz}

        \caption{Follow-up test case after LC at node 2. The stabilizer element becomes
        $K_1' = Y_1 Y_2$ and the measurement circuit changes accordingly.
        MR modifications are marked with dashed red lines.}
        \label{fig:lc-example-followup}
    \end{subfigure}

    \caption{Example of a local-complementation-based metamorphic relation for graph states.}
    \label{fig:LC_implementation}
\end{figure*}

\paragraph{LC} A second relation is based on local complementation, a graph operation
that toggles all edges among the neighbors of a chosen vertex \cite{vandennest_lc_2004}. Although
this can change the structure of the graph considerably, the corresponding
graph states differ only by single-qubit Clifford operations (that is, a concatenation
of the single-qubit gates $H$ and $S$) on the chosen vertex
and its neighbors (\cref{fig:LC_implementation}). A correct program must therefore produce a follow-up
state that is related to the source state in this simple way. The expected
measurements of the follow-up state are obtained by transforming those of
the source state with the same single-qubit operations, so they remain
deterministic, keep their weight, and have the same expected outcomes.
Since the program has to prepare a different set of edges, this relation can reveal
faults in how the program translates the edges of a graph into a circuit,
even though the expected state changes only locally.

\paragraph{CNOT} A third relation is based on $\mathrm{CNOT}$ gates. Applying a
$\mathrm{CNOT}$ gate to a graph state, followed by a $Z$ gate on the
control qubit if control and target are connected, yields another graph
state, whose graph is obtained by toggling the edges between the control
vertex and the neighbors of the target vertex \cite{Gachechiladze_2017UnitaryGraph}. In contrast to the previous
relations, source and follow-up state are related by an entangling
two-qubit operation rather than by single-qubit operations or a
relabeling, so this relation connects graph states with genuinely different
entanglement. To verify the follow-up state, we undo this operation before
the measurement, i.e., we apply the $Z$ gate (if required) and the
$\mathrm{CNOT}$ gate again, which maps a correct follow-up state back to
the source state, and then measure the original stabilizer elements. The
measured Pauli strings, and hence their weights, are therefore unchanged,
and their outcomes remain deterministic with the same expected values. This
relation differs from the two first ones as it requires non-local gates to be performed
in the measurement process, but enhances the variety of follow-up states.

\begin{figure}[t]
    \centering
    \resizebox{0.8\columnwidth}{!}{%
    \begin{tikzpicture}[
        v/.style={circle, draw, fill=white, minimum size=3.6mm, inner sep=0pt, font=\scriptsize},
        e/.style={thick},
        op/.style={-{Stealth[length=2mm]}, semithick},
        oplabel/.style={font=\footnotesize, anchor=south, inner sep=2pt},
        gname/.style={font=\small, anchor=north}
    ]
        \def\gpanel#1#2{%
            \begin{scope}[shift={(#1,0)}]
                \node[v] (#2-1) at (0,0.75)    {1};
                \node[v] (#2-2) at (0.75,0.75) {2};
                \node[v] (#2-3) at (0.75,0)    {3};
                \node[v] (#2-4) at (0,0)       {4};
            \end{scope}}

        \gpanel{0}{a}
        \draw[e] (a-1)--(a-2) (a-2)--(a-3) (a-3)--(a-4) (a-4)--(a-1) (a-1)--(a-3);
        \node[gname] at (0.375,-0.3) {$G_0$};

        \gpanel{2.8}{b}
        \draw[e] (b-1)--(b-2) (b-2)--(b-3) (b-3)--(b-4) (b-4)--(b-1) (b-2)--(b-4);
        \node[gname] at (3.175,-0.3) {$G_1$};

        \gpanel{5.6}{c}
        \draw[e] (c-1)--(c-2) (c-2)--(c-3) (c-1)--(c-3) (c-2)--(c-4);
        \node[gname] at (5.975,-0.3) {$G_2$};

        \gpanel{8.4}{d}
        \draw[e] (d-1)--(d-2) (d-1)--(d-3) (d-2)--(d-4);
        \node[gname] at (8.775,-0.3) {$G_3$};

        \foreach \x/\lab in {0/{\pi}, 2.8/{\mathrm{LC}_2}, 5.6/{\mathrm{CNOT}_{34}}}{
            \draw[op] (\x+1.4,0.375) -- node[oplabel] {$\lab$} (\x+2.15,0.375);
        }
    \end{tikzpicture}%
    }
    \caption{A chain of graph transformations: $G_1 = \pi(G_0)$ with the qubit
    permutation $\pi = (1\,2\,3\,4)$; $G_2 = \mathrm{LC}_2(G_1)$, local
    complementation at vertex 2; and $G_3 = \mathrm{CNOT}_{34}(G_2)$, a CNOT
    with control 3 and target 4.}
    \label{fig:graph-transformations}
\end{figure}
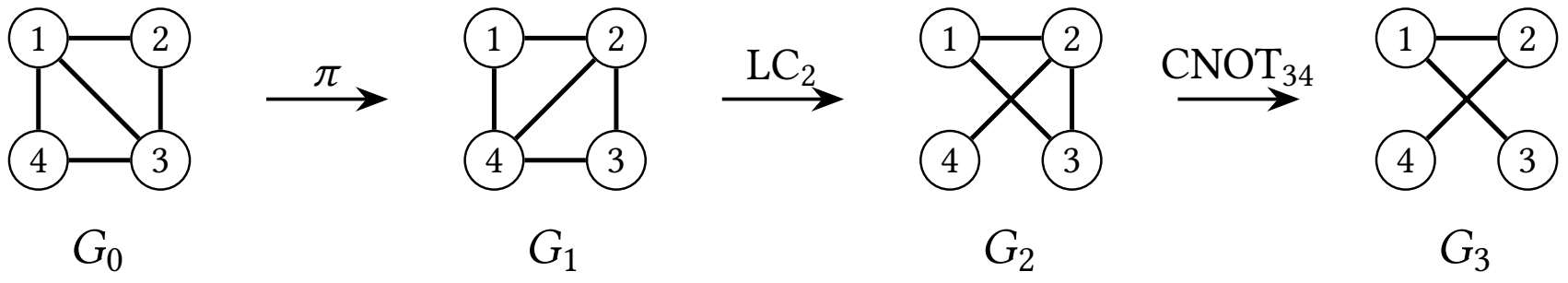

Since all metamorphic relations map graph states to graph states, they can be
composed freely (\cref{fig:graph-transformations}). How each relation transforms
the input state follows directly from its definition above. The transformation
of the measurement procedure, however, depends on which relations are combined.
If only \emph{PERM} and/or \emph{LC} are used, it suffices to update the
single-qubit Pauli operators in each Pauli string of the final measurement.
If \emph{CNOT} is also applied, this is no longer sufficient. Instead, we insert
a pre-measurement layer directly before the stabilizer measurement. This layer
undoes the $\mathrm{CNOT}$ (and $Z$) gates as well as the local gates introduced by \emph{LC}, while
\emph{PERM} is handled by relabeling the qubits of the measurement circuit and
of the final measurements.

\section{Experimental evaluation}
In this section, we present the evaluation results for the methods introduced in Section~\ref{sec:approach}. We consider several variants of the metamorphic testing approach, using different combinations of the metamorphic relations \textit{LC}, \textit{CNOT}, and \textit{PERM}. In addition, we include a baseline method that does not generate follow-up test cases. Based on this experimental setup, we address the following research questions.

\subsection{Research questions}

\medskip

\noindent\textbf{(RQ1) Efficiency.} How does the computational cost of the evaluated testing methodologies compare?

\begin{itemize}
\item[(a)] What is the runtime for the different testing methodologies and across different distributions of MRs?

\item[(b)] What is the time to first mutation detection for each testing methodology?

\end{itemize}

\noindent\textbf{(RQ2) Effectiveness.} How does the fault detection effectiveness of the evaluated testing methodologies compare?

\begin{itemize}
\item[(a)] What is the fault detection rate for the different testing methodologies and across different distributions of MRs?

\item[(b)] Can the proposed Metamorphic Testing approach detect previously unknown bugs in the current Qiskit compiler?

\end{itemize}

\subsection{Experimental setup}

As the function under test, we use the Qiskit function
\texttt{GraphStateGate()}~\cite{ibm_graphstategate}. Given the  graph~$G$,
it returns a quantum circuit that prepares the corresponding
graph state~$\ket{G}$. As inputs, we generate random Erd\H{o}s--R\'enyi graphs
with 24 nodes, in which each pair of nodes is connected independently with
probability~$p$~\cite{erdos59a}. We choose $p = 0.5$ because then the expected
number of edges is the same before and after applying \emph{LC} and
\emph{CNOT}. This prevents the metamorphic relations from affecting the runtime
of simulating the circuits. We further postselect graphs whose stabilizer
group contains an element of weight at most $w_{\max}=4$. This is the lowest
weight bound for which our sampler found such graphs within 1000 attempts.

Given the relatively low complexity of the \texttt{GraphStateGate()} function, we do not expect to identify faults in its current Qiskit implementation. Therefore, we introduce a set of mutation operators to inject artificial faults into the generated circuits. The selection of these mutation operators is not arbitrary, as it is based on mutation operators previously considered in the literature \cite{mendiluze_muskit_2021}, together with an evaluation of different fault types and their applicability to graph-state circuit generation. The resulting classes of mutation operators are as follows:

\begin{itemize}
    \item \textbf{\texttt{Insert random gate}}: insert a random unitary gate somewhere in the created circuit, from a gate set \(\{X, Y, Z, H, CNOT, CZ\}\). 
    \item  \textbf{\texttt{Remove random gate}:} remove a random gate from the created circuit.
    \item \textbf{\texttt{Reorder gates}:} reorders a pair of two gates from the created circuit. Note that these gates have to be different. Otherwise, for graph state creation, it would lead to an equivalent mutant.
    \item \textbf{\texttt{Change qubits}:} changes the qubit or qubits associated with a gate (e.g. \textit{CZ(1,3)} instead of \textit{CZ(1,2)}). 
    \item \textbf{\texttt{Incorrect qubit initialization}:} incorrectly initializes a random qubit by introducing a unitary gate at the beginning of the circuit for said qubit. 
\end{itemize}

Once a type of mutation operator is selected for a so-called \textit{trial}, all executions belonging to said trial share a common seed to ensure that the mutation operator affects the same specific qubits in approximately the same layer. 

Additionally, we consider four variants of the metamorphic testing approach, defined by different probability distributions over the metamorphic relations LC, PERM, and CNOT. We denote these probabilities by \(p_{LC}\), \(p_{PERM}\), and \(p_{CNOT}\), respectively. The probability distributions used for each variant are shown in Table~\ref{tab:MR_probabilities}.

\begin{table}[t]
\centering
\caption{Probability of selecting each metamorphic relation in each testing variant.}
\label{tab:MR_probabilities}

\begin{tabular}{lcccc}
\toprule
 & \textit{LC\_only} & \textit{LC\_PERM} & \textit{All} & \textit{CNOT\_PERM} \\
\midrule
$p_{LC}$   & 1.0 & 0.8 & 0.4 & 0.0 \\
$p_{PERM}$    & 0.0 & 0.2 & 0.2 & 0.2 \\
$p_{CNOT}$ & 0.0 & 0.0 & 0.4 & 0.8 \\
\bottomrule
\end{tabular}
\end{table}

We selected these MR distributions to represent the main properties of the proposed testing approach that we aim to investigate. In particular, they allow us to evaluate how the method performs when restricted to basic relations such as LC, whether incorporating PERM relations provides a meaningful benefit, and whether the additional cost of CNOT-based MRs is justified by the effectiveness they provide. While many additional MR distributions could be evaluated, we found them to be less relevant to the specific properties investigated in this study based on our experience with the approach. We always chain five random metamorphic relations to generate the next test case to ensure a broad coverage of graphs.

\subsection{Data Collection and Measurement Setup}

For \textbf{RQ1a}, we record the mean runtime (s) of different trials for both the baseline and the metamorphic testing method, considering the probabilities expressed in Table~\ref{tab:MR_probabilities}. We record the runtime for four different sub-processes inside each execution: \textit{Graph and low-weight stabilizer generation}, \textit{application of MRs}, \textit{optimization of the measurement circuit}, and \textit{program execution}. We perform 500 trials, with each trial consisting of 204 test cases. For the Metamorphic Testing method, we provide four source test cases and derive 50 test cases from each, through steps of applying five metamorphic relations.  

For \textbf{RQ1b}, we collect the runtime until fault detection (s) across \(2500\) different trials across all classes of mutation operators. That is, the recorded runtime it takes for each method to identify the mutant in each trial. We also record the number of test cases until mutation detection. We categorize the results by mutant operator class, executing \(500\) trials for each. Each trial consists of the same number of executions, source test cases and MRs as in RQ1a. 

For \textbf{RQ2a}, we record the fault detection rate across \(2500\) trials and categorize the results by mutant operator class. The number of executions per trial is the same as in RQ1b.

As for \textbf{RQ2b}, we test the Qiskit transpiler instead of
\texttt{GraphStateGate()}. The program under test is the graph-state preparation
circuit compiled by the transpiler, and no mutations are injected. We reduced
the number of qubits to $12$ to speed up the testing. We generate
10~source test cases as described above and derive 100~chained follow-ups from
each, applying five MRs per follow-up with the \emph{LC\_PERM} variant.
Each of the resulting 1010~test cases is compiled for a 14-qubit
device with a randomly sampled transpiler configuration. We sample the
optimization level (0--3), one of five basis gate sets, the topology
(\textit{all-to-all},\textit{ line}, \textit{ring}, \textit{grid}, \textit{heavy-hex}), whether coupling-map edges are directed, the layout method (\textit{trivial}, \textit{dense}, \textit{sabre}), the routing method (\textit{basic}, \textit{sabre}, \textit{lookahead}), and the transpiler seed. Only the state preparation
is compiled. The measurement circuit and the stabilizer measurements are
appended on the physical qubits to which the compiler mapped the logical
qubits, and every stabilizer must yield $+1$ in all of 32~shots. To separate
compiler bugs from errors in the test itself, each test case is first executed
without compilation. We record the outcome of every test case as pass, fail
(a violated stabilizer), compiler error, or timeout (compilation not finished
within 30\,s).

\subsection{Results} 

\begin{figure}[t]
\centering
\begin{tikzpicture}

\pgfplotsset{
    runtime split/.style={
        width=0.88\linewidth,
        scale only axis,
        ybar stacked,
        bar width=36pt,
        xmin=0.5,
        xmax=5.5,
        enlargelimits=false,
        axis x line*=bottom,
        axis y line*=left,
        tick align=outside,
        ymajorgrids=true,
        grid style={gray!20},
        clip=true,
        clip mode=global,
        tick label style={font=\small}
    }
}

\begin{axis}[
    runtime split,
    name=lower,
    height=2.6cm,
    ymin=0,
    ymax=4,
    ytick={0,1,2,3},
    xtick={1,2,3,4,5},
    xticklabels={
        Baseline,
        {LC\\only},
        {LC\\PERM},
        {All},
        {CNOT\\PERM}
    },
    xticklabel style={align=center,font=\small}
]
\addplot[fill=generation,draw=white,line width=0.5pt] coordinates {(1,62.9391615860) (2,1.2215092453) (3,1.2054824780) (4,1.2031746322) (5,1.2603110168)};
\addplot[fill=mr,draw=white,line width=0.5pt] coordinates {(1,0.0000000000) (2,0.0592582109) (3,0.0568785388) (4,0.0790768963) (5,0.0610501301)};
\addplot[fill=optimization,draw=white,line width=0.5pt] coordinates {(1,0.0000000000) (2,0.0000000000) (3,0.0000000000) (4,0.4449341585) (5,0.3060716947)};
\addplot[fill=execution,draw=white,line width=0.5pt] coordinates {(1,1.0708695521) (2,0.9926773531) (3,0.9994267130) (4,1.6112448212) (5,1.3332489311)};
\node[anchor=south,font=\small,yshift=3pt] at (axis cs:1,64.0100311381) {64.02\,s};
\node[anchor=south,font=\small,yshift=3pt] at (axis cs:2,2.2734448094) {2.28\,s};
\node[anchor=south,font=\small,yshift=3pt] at (axis cs:3,2.2617877298) {2.27\,s};
\node[anchor=south,font=\small,yshift=3pt] at (axis cs:4,3.3384305083) {3.48\,s};
\node[anchor=south,font=\small,yshift=3pt] at (axis cs:5,2.9606817727) {3.04\,s};
\end{axis}

\begin{axis}[
    runtime split,
    name=upper,
    at={(lower.north west)},
    anchor=south west,
    yshift=12pt,
    height=1.5cm,
    ymin=62.5,
    ymax=65,
    ytick={63,64,65},
    xtick=\empty,
    axis x line=none,
    legend columns=2,
    legend style={
        at={(0.5,1.15)},
        anchor=south,
        draw=none,
        fill=none,
        font=\footnotesize
    }
]
\addplot[fill=generation,draw=white,line width=0.5pt] coordinates {(1,62.9391615860) (2,1.2215092453) (3,1.2054824780) (4,1.2031746322) (5,1.2603110168)};
\addlegendentry{Graph + stabilizer generation}
\addplot[fill=mr,draw=white,line width=0.5pt] coordinates {(1,0.0000000000) (2,0.0592582109) (3,0.0568785388) (4,0.0790768963) (5,0.0610501301)};
\addlegendentry{MR application}
\addplot[fill=optimization,draw=white,line width=0.5pt] coordinates {(1,0.0000000000) (2,0.0000000000) (3,0.0000000000) (4,0.4449341585) (5,0.3060716947)};
\addlegendentry{Measurement circuit optimization}
\addplot[fill=execution,draw=white,line width=0.5pt] coordinates {(1,1.0708695521) (2,0.9926773531) (3,0.9994267130) (4,1.6112448212) (5,1.3332489311)};
\addlegendentry{Execution}
\node[anchor=south,font=\small,yshift=3pt] at (axis cs:1,64.0100311381) {64.02\,s};
\node[anchor=south,font=\small,yshift=3pt] at (axis cs:2,2.2734448094) {2.28\,s};
\node[anchor=south,font=\small,yshift=3pt] at (axis cs:3,2.2617877298) {2.27\,s};
\node[anchor=south,font=\small,yshift=3pt] at (axis cs:4,3.3384305083) {3.48\,s};
\node[anchor=south,font=\small,yshift=3pt] at (axis cs:5,2.9606817727) {3.04\,s};
\end{axis}

\coordinate (baselineLower) at ($(lower.north west)!0.1!(lower.north east)$);
\coordinate (baselineUpper) at ($(upper.south west)!0.1!(upper.south east)$);

\begin{scope}
    \clip
        ([xshift=-18pt]baselineLower)
        rectangle
        ([xshift=18pt]baselineUpper);

    \foreach \x in {-40,-35,...,35}{
        \draw[
            pythonblue,
            line width=1pt,
            dotted
        ]
        ([xshift=\x pt,yshift=-2.3pt]baselineLower)
        --
        ++(0pt,28pt);
    }
\end{scope}

\draw[thick]
    ([xshift=-3pt,yshift=-2pt]lower.north west)
    -- ([xshift=3pt,yshift=2pt]lower.north west);

\draw[thick]
    ([xshift=-3pt,yshift=-2pt]upper.south west)
    -- ([xshift=3pt,yshift=2pt]upper.south west);

\node[rotate=90,anchor=south]
    at ([xshift=-1cm,yshift=2.3cm]lower.south west)
    {Mean runtime per trial (s)};

\end{tikzpicture}
\caption{Mean runtime components across 500 trials. Numeric labels show the mean recorded total runtime. The y-axis is split to accommodate the substantially larger runtime of the baseline while preserving the visibility of the remaining approaches.}
\label{fig:runtime}
\end{figure}
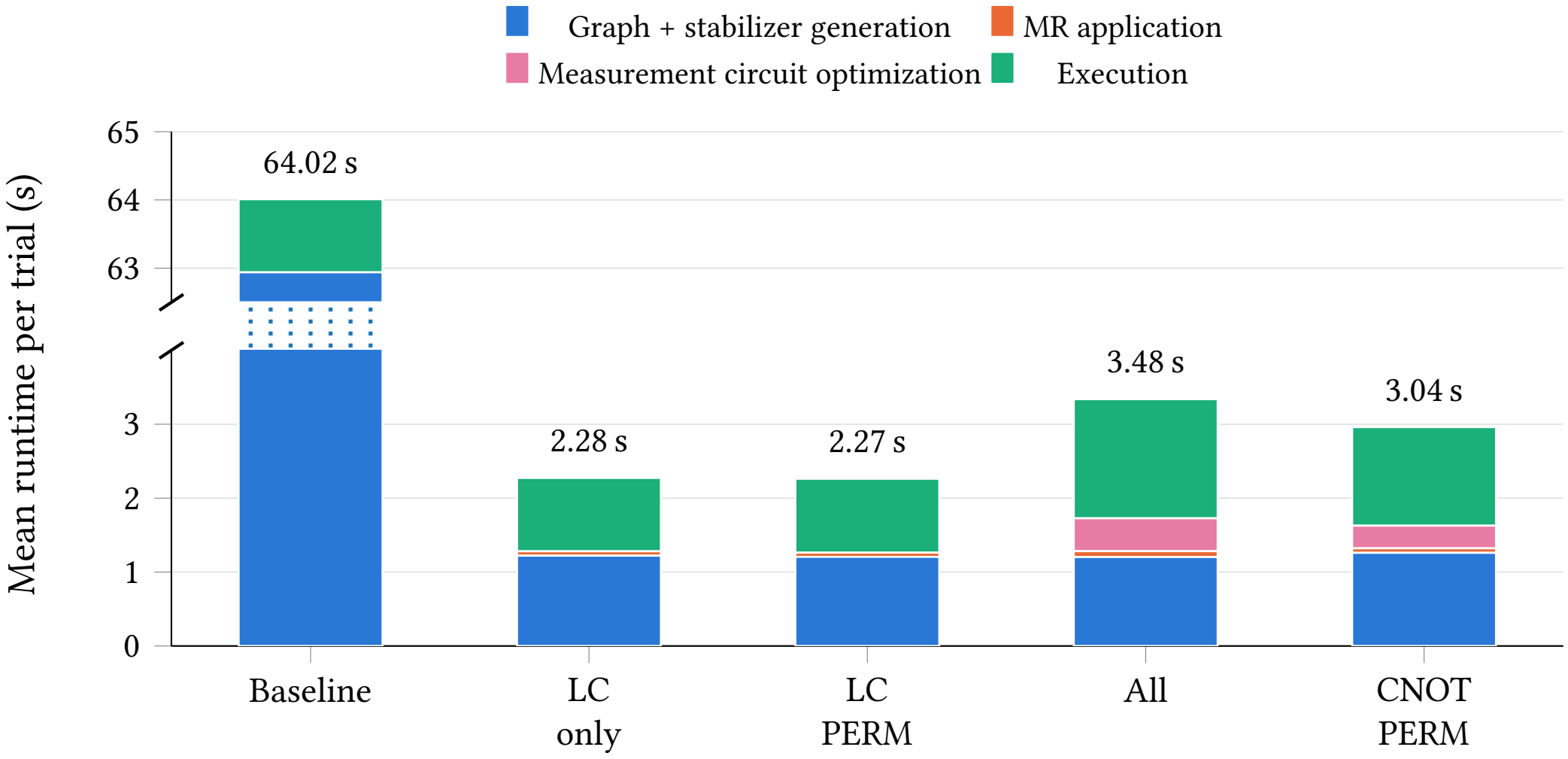

\textbf{RQ1a}. Figure \ref{fig:runtime} shows the mean total runtime per trial, measured in seconds, for the baseline and the metamorphic testing approaches under different combinations of MRs. The baseline exhibits a substantially higher runtime, with a mean of 64.02 s. In contrast, \textit{LC\_only} and \textit{LC\_PERM} have mean runtimes of approximately 2.3 s. Approaches incorporating CNOT MRs show slightly higher runtimes, with 3.48 s for \textit{All} and 3.04 s for \textit{CNOT\_PERM}.

\textbf{RQ1b}. Figure \ref{fig:time_until_detection} shows the time to first mutation detection per trial, measured in seconds on a logarithmic scale, for all methodologies. All metamorphic testing approaches detect mutations considerably faster than the baseline, which takes a median of 1.7\,s. \textit{LC\_PERM}, \textit{All}, and \textit{CNOT\_PERM} reach a median of about 0.27\,s, roughly six times faster than the baseline. \textit{LC\_only}, which considers only Local Complementation (LC) as an MR, is slower than the other variants, with a median of 0.38\,s, but still about 4.5 times faster than the baseline.

Figure \ref{fig:test_cases_until_detection} shows the number of test cases required by each methodology to detect a mutation in a trial. The \textit{LC\_only} variant of the metamorphic testing approach requires substantially more test cases on average than the other methodologies. In contrast, the baseline and the remaining metamorphic testing variants require between 6 and 10 test cases on average before detecting a mutation. The approaches incorporating CNOT MRs, \textit{All} and \textit{CNOT\_PERM}, require a mean of 7.4 and 6.6 test cases, respectively, compared with 8.3 test cases for the baseline.

\begin{figure*}[t]
\centering

\begin{minipage}[t]{0.49\textwidth}
\centering

\begin{tikzpicture}
\begin{axis}[
    width=\linewidth,
    height=5.5cm,
    yminorgrids=true,
    minor grid style={gray!10},
    boxplot/draw direction=y,
    boxplot/box extend=0.5,
    ymode=log,
    ymin=0.007,
    ymax=100,
    xmin=0.5,
    xmax=5.5,
    ylabel={Runtime to first detection (s)},
    xtick={1,2,3,4,5},
    xticklabels={
        Baseline,
        {LC\\only},
        {LC\\PERM},
        {All},
        {CNOT\\PERM}
    },
    xticklabel style={align=center,font=\scriptsize},
    tick align=outside,
    clip=false
]
\addplot[boxplot prepared={draw position=1,lower whisker=0.0211776521,lower quartile=0.6839824191,median=1.7144706056,upper quartile=3.4676934939,upper whisker=7.6205181982},fill=boxbaseline,draw=black!70,mark=none] coordinates {(0,7.6593289888) (0,7.6644498976) (0,7.6754130875) (0,7.6905849052) (0,7.7071171149) (0,7.7312008808) (0,7.7317185919) (0,7.7564594892) (0,7.7572028693) (0,7.7615209084) (0,7.7946302239) (0,7.8236601932) (0,7.8419808496) (0,7.8453527973) (0,7.8961537257) (0,7.9264413449) (0,7.9393658575) (0,7.9452447447) (0,7.9471549930) (0,7.9519335229) (0,7.9547330611) (0,8.0030451890) (0,8.0058530185) (0,8.0856336276) (0,8.0981859171) (0,8.1454524605) (0,8.1751596690) (0,8.2172264245) (0,8.2174271685) (0,8.2579420372) (0,8.2982597530) (0,8.3013186010) (0,8.3234084668) (0,8.3263193783) (0,8.3282843339) (0,8.3644719643) (0,8.3751368145) (0,8.3758821923) (0,8.4153151887) (0,8.4333949778) (0,8.4663692189) (0,8.4919621730) (0,8.5260368488) (0,8.5406058715) (0,8.5683142734) (0,8.5712355736) (0,8.6007512624) (0,8.6225571083) (0,8.6226125329) (0,8.7550329375) (0,8.7970095442) (0,8.8313774518) (0,8.8552122258) (0,8.8710497415) (0,8.9892766268) (0,9.0090492801) (0,9.0342970770) (0,9.0877690369) (0,9.1077751408) (0,9.2577528229) (0,9.2893075722) (0,9.3106755624) (0,9.3401659094) (0,9.3897407388) (0,9.5542391145) (0,9.6126067527) (0,9.7144266448) (0,9.7673873107) (0,9.7868114843) (0,9.8186027855) (0,9.8615033003) (0,9.9655042919) (0,10.0606702014) (0,10.1121149121) (0,10.1557958710) (0,10.1974454881) (0,10.2481709740) (0,10.3195616221) (0,10.3425304901) (0,10.3598789934) (0,10.4658392628) (0,10.6013286745) (0,10.6263955033) (0,10.6383025122) (0,10.7726401035) (0,10.7790104509) (0,10.8179834259) (0,11.1323227994) (0,11.1908278067) (0,11.2403047683) (0,11.3653982179) (0,11.4132882149) (0,11.4249643513) (0,11.4343192235) (0,11.4611046088) (0,11.5066396280) (0,11.5201072746) (0,11.8176381662) (0,12.2092958021) (0,12.2884618249) (0,12.3822259428) (0,12.3953814725) (0,12.4470393604) (0,12.5897052409) (0,12.7746205963) (0,12.9126811812) (0,12.9658820031) (0,13.0765171777) (0,13.1017949299) (0,13.2946223905) (0,13.3122112132) (0,13.4652587639) (0,14.1665064602) (0,14.2829473519) (0,14.6080905329) (0,14.7985609884) (0,15.3148333575) (0,15.3885328323) (0,15.7027776763) (0,15.9409506179) (0,16.0722114069) (0,16.2813392577) (0,16.3268982035) (0,16.5021559908) (0,16.9294015577) (0,16.9317581919) (0,17.3389644413) (0,17.9074145602) (0,18.0425264530) (0,20.1284377198) (0,20.4454147888) (0,21.7324579065) (0,21.7950836155) (0,22.2603700906) (0,22.5262115481) (0,22.9321629505) (0,22.9851984766) (0,23.1872723103) (0,27.8991057589) (0,32.0364218433) (0,32.8765038261) (0,33.3112693864)};
\node[anchor=south,font=\scriptsize,yshift=-7pt] at (axis cs:1,38.3079597944) {$n=2458$};
\addplot[boxplot prepared={draw position=2,lower whisker=0.0209933899,lower quartile=0.1702440681,median=0.3755438637,upper quartile=0.7785526837,upper whisker=1.6874636745},fill=boxlconly,draw=black!70,mark=none] coordinates {(0,1.6933702109) (0,1.7004588554) (0,1.7016477017) (0,1.7059771870) (0,1.7110395534) (0,1.7199509968) (0,1.7255668209) (0,1.7274854106) (0,1.7281808138) (0,1.7373112941) (0,1.7401917512) (0,1.7531178682) (0,1.7540611704) (0,1.7549290161) (0,1.7551600644) (0,1.7583672698) (0,1.7605110435) (0,1.7638568024) (0,1.7651839994) (0,1.7706890635) (0,1.7801522959) (0,1.7802983199) (0,1.7863741270) (0,1.8019368402) (0,1.8047099463) (0,1.8056095473) (0,1.8062294882) (0,1.8075890988) (0,1.8115939284) (0,1.8133270552) (0,1.8167424055) (0,1.8288261662) (0,1.8296142546) (0,1.8304380323) (0,1.8315847691) (0,1.8433686509) (0,1.8440674772) (0,1.8442191174) (0,1.8482035699) (0,1.8530527465) (0,1.8557733572) (0,1.8561016854) (0,1.8677900245) (0,1.8704971089) (0,1.8718755718) (0,1.8838085516) (0,1.8955787013) (0,1.9017752728) (0,1.9047938760) (0,1.9154255313) (0,1.9168842128) (0,1.9207871330) (0,1.9547331887) (0,1.9578844616) (0,1.9618201621) (0,1.9654020786) (0,1.9977122692) (0,2.0062626312) (0,2.0083901689) (0,2.0296897471) (0,2.0402891093) (0,2.0601843542) (0,2.0629430348) (0,2.0657643420) (0,2.0659552016) (0,2.0723769222) (0,2.1022242429) (0,2.1131703376) (0,2.1192447811) (0,2.1204106230) (0,2.1287383509) (0,2.1292457038) (0,2.1323458708) (0,2.1343372609) (0,2.1445738736) (0,2.1698358601) (0,2.1773133678) (0,2.1944211840) (0,2.2041469836) (0,2.2041847608) (0,2.2164701070) (0,2.2260489599) (0,2.2436147877) (0,2.2511988508) (0,2.2686210100) (0,2.2741399421) (0,2.2783888464) (0,2.2947222164) (0,2.3175817013) (0,2.3183098715) (0,2.3387142061) (0,2.3466714169) (0,2.3468116026) (0,2.3472221894) (0,2.3477866021) (0,2.3640173678) (0,2.3934574577) (0,2.4058037885) (0,2.4229626867) (0,2.4396350519) (0,2.4432231335) (0,2.4605249446) (0,2.4992061732) (0,2.5137057607) (0,2.5389223997) (0,2.5429363891) (0,2.5770459643) (0,2.6000024972) (0,2.6072043295) (0,2.6116987916) (0,2.6336661389) (0,2.6367114317) (0,2.6428250147) (0,2.6431497452) (0,2.6582731029) (0,2.6652786057) (0,2.6965288743) (0,2.8492050429) (0,2.8845809549) (0,2.8865609376) (0,2.9128129261) (0,2.9754695289) (0,2.9811087563) (0,2.9937820123) (0,3.0406762485) (0,3.0766755857) (0,3.1230590697) (0,3.1539350206) (0,3.2799568027) (0,3.4708690974) (0,3.8825445715) (0,4.3637898038) (0,4.3703402900) (0,4.3733087345)};
\node[anchor=south,font=\scriptsize,yshift=-7pt] at (axis cs:2,5.0293050447) {$n=2103$};
\addplot[boxplot prepared={draw position=3,lower whisker=0.0214536141,lower quartile=0.1515551457,median=0.2767760415,upper quartile=0.4835501161,upper whisker=0.9709355654},fill=boxlcshuffle,draw=black!70,mark=none] coordinates {(0,0.9819699556) (0,0.9824662202) (0,0.9825064209) (0,0.9833906789) (0,0.9859722422) (0,0.9893195345) (0,0.9966633986) (0,0.9983816957) (0,1.0004055740) (0,1.0017626784) (0,1.0188781179) (0,1.0197205008) (0,1.0198304551) (0,1.0202820485) (0,1.0217496920) (0,1.0334325421) (0,1.0345042737) (0,1.0352368432) (0,1.0371688651) (0,1.0401482955) (0,1.0424725041) (0,1.0509845042) (0,1.0513776122) (0,1.0582267726) (0,1.0584496974) (0,1.0604061889) (0,1.0613225894) (0,1.0631217519) (0,1.0635958610) (0,1.0641460533) (0,1.0642709967) (0,1.0724804879) (0,1.0765944545) (0,1.0793933722) (0,1.0801947978) (0,1.0822364641) (0,1.0847988294) (0,1.0849608639) (0,1.0963399413) (0,1.1007252487) (0,1.1042191992) (0,1.1138753730) (0,1.1148096300) (0,1.1167809570) (0,1.1174083911) (0,1.1230381359) (0,1.1249143588) (0,1.1271083136) (0,1.1301562793) (0,1.1317634049) (0,1.1367878828) (0,1.1398129466) (0,1.1489966223) (0,1.1549652587) (0,1.1581765686) (0,1.1642030848) (0,1.1791170416) (0,1.1900368498) (0,1.2011994342) (0,1.2058671957) (0,1.2145297402) (0,1.2172951750) (0,1.2210541647) (0,1.2290061684) (0,1.2335178561) (0,1.2371335737) (0,1.2439579468) (0,1.2441963721) (0,1.2483390770) (0,1.2542186503) (0,1.2572007680) (0,1.2741076080) (0,1.2745890855) (0,1.2800600117) (0,1.2857878332) (0,1.2870296321) (0,1.2939381176) (0,1.3071888746) (0,1.3087682861) (0,1.3119578620) (0,1.3169089777) (0,1.3196682422) (0,1.3329420320) (0,1.3567486650) (0,1.3568705730) (0,1.3696152871) (0,1.3709712829) (0,1.3775669055) (0,1.3818378986) (0,1.3924926699) (0,1.3937367005) (0,1.3987458553) (0,1.4010698900) (0,1.4063386992) (0,1.4264594240) (0,1.4381897249) (0,1.4515690985) (0,1.4602460477) (0,1.4810087655) (0,1.5026634699) (0,1.5046528389) (0,1.5110073800) (0,1.5121510897) (0,1.5178130402) (0,1.5180103460) (0,1.5214160979) (0,1.5274592345) (0,1.5548369677) (0,1.5645795078) (0,1.5718857578) (0,1.6591333575) (0,1.6836388332) (0,1.6935363784) (0,1.6994010019) (0,1.7142631991) (0,1.7162995394) (0,1.7320645568) (0,1.7700603229) (0,1.7967942120) (0,1.8196941831) (0,1.8666210894) (0,1.9096951310) (0,1.9212757240) (0,2.0144956948) (0,2.8822346092)};
\node[anchor=south,font=\scriptsize,yshift=-7pt] at (axis cs:3,3.3145698006) {$n=2460$};
\addplot[boxplot prepared={draw position=4,lower whisker=0.0211996289,lower quartile=0.1483799771,median=0.2704990781,upper quartile=0.4707031124,upper whisker=0.9537727309},fill=boxall,draw=black!70,mark=none] coordinates {(0,0.9560736706) (0,0.9563399148) (0,0.9585443409) (0,0.9608512891) (0,0.9611600877) (0,0.9634557511) (0,0.9642815304) (0,0.9668687282) (0,0.9681999029) (0,0.9682332880) (0,0.9739794361) (0,0.9752917548) (0,0.9757492831) (0,0.9779855476) (0,0.9814157891) (0,0.9826311185) (0,0.9910751430) (0,0.9991267286) (0,0.9991760650) (0,0.9994913426) (0,0.9998346271) (0,1.0123721608) (0,1.0137599718) (0,1.0184986847) (0,1.0207759421) (0,1.0226635910) (0,1.0268397504) (0,1.0285819431) (0,1.0297887207) (0,1.0302135230) (0,1.0309957680) (0,1.0310524195) (0,1.0338729643) (0,1.0345606257) (0,1.0348281434) (0,1.0351293182) (0,1.0385703903) (0,1.0387264504) (0,1.0412429061) (0,1.0426133689) (0,1.0490546969) (0,1.0518956522) (0,1.0527350670) (0,1.0543766229) (0,1.0606356540) (0,1.0707450991) (0,1.0722155056) (0,1.0835935054) (0,1.0861078012) (0,1.0966146449) (0,1.0990788185) (0,1.0991731083) (0,1.1014844421) (0,1.1030135427) (0,1.1062648250) (0,1.1072584454) (0,1.1082204201) (0,1.1188704770) (0,1.1196298169) (0,1.1209704978) (0,1.1212036132) (0,1.1255141932) (0,1.1283036834) (0,1.1301755328) (0,1.1308022582) (0,1.1360030926) (0,1.1367003715) (0,1.1481246583) (0,1.1483422508) (0,1.1538893101) (0,1.1675746790) (0,1.1716614340) (0,1.1737436154) (0,1.1751060272) (0,1.1753227527) (0,1.1911736894) (0,1.2002095808) (0,1.2114730089) (0,1.2118624940) (0,1.2246234147) (0,1.2255248909) (0,1.2377943816) (0,1.2385994950) (0,1.2473249862) (0,1.2615345158) (0,1.2724093883) (0,1.2790843549) (0,1.2832136345) (0,1.2846883335) (0,1.2889537443) (0,1.2921213736) (0,1.2941233222) (0,1.2960523521) (0,1.3043026554) (0,1.3211104595) (0,1.3235290884) (0,1.3255708176) (0,1.3348668693) (0,1.3358615413) (0,1.3447655886) (0,1.3511055787) (0,1.3705888826) (0,1.3738666940) (0,1.3768903255) (0,1.3808306193) (0,1.3838425442) (0,1.4001601741) (0,1.4149455111) (0,1.4161023216) (0,1.4290896340) (0,1.4372601828) (0,1.4397528775) (0,1.4428499455) (0,1.4614539859) (0,1.4624343209) (0,1.4668232349) (0,1.4694240729) (0,1.4699991376) (0,1.4715835727) (0,1.4790918594) (0,1.4837875445) (0,1.4851433919) (0,1.5199066743) (0,1.5524192238) (0,1.5675374009) (0,1.6596567316) (0,1.7767617803) (0,1.7795055523) (0,1.7980732189) (0,1.8362806891) (0,1.9140662039) (0,2.1149523819) (0,2.1263901927) (0,2.2856668199) (0,2.4472662490) (0,2.4669399702) (0,2.5116606615) (0,2.5129892624) (0,3.4963846065)};
\node[anchor=south,font=\scriptsize,yshift=-7pt] at (axis cs:4,4.0208422974) {$n=2460$};
\addplot[boxplot prepared={draw position=5,lower whisker=0.0217329219,lower quartile=0.1434169698,median=0.2687637138,upper quartile=0.4802845791,upper whisker=0.9847184392},fill=boxcnot,draw=black!70,mark=none] coordinates {(0,0.9902146531) (0,0.9907889380) (0,0.9908799296) (0,0.9921400268) (0,0.9953003989) (0,0.9997474509) (0,1.0017334048) (0,1.0021392575) (0,1.0025611562) (0,1.0082730178) (0,1.0085987463) (0,1.0087418368) (0,1.0120639259) (0,1.0184009240) (0,1.0193828403) (0,1.0234140872) (0,1.0248344913) (0,1.0262196320) (0,1.0281422958) (0,1.0335226685) (0,1.0354136329) (0,1.0484230670) (0,1.0561617324) (0,1.0571375913) (0,1.0716602891) (0,1.0745474130) (0,1.0768073162) (0,1.0768507572) (0,1.0769600002) (0,1.0784787743) (0,1.0792412187) (0,1.0822069349) (0,1.0824852476) (0,1.0853619722) (0,1.0948297274) (0,1.0992090262) (0,1.1002611588) (0,1.1030166501) (0,1.1039868521) (0,1.1051994378) (0,1.1099396925) (0,1.1127877510) (0,1.1153257729) (0,1.1160939126) (0,1.1183581911) (0,1.1195083703) (0,1.1270742407) (0,1.1290951322) (0,1.1294103682) (0,1.1341769150) (0,1.1364658263) (0,1.1511667310) (0,1.1562194272) (0,1.1585090775) (0,1.1603494047) (0,1.1631986629) (0,1.1671733558) (0,1.1706752426) (0,1.1707350928) (0,1.1710435008) (0,1.1771349802) (0,1.1799435974) (0,1.1812847783) (0,1.1898155480) (0,1.1899999615) (0,1.1916641942) (0,1.2027980459) (0,1.2044986961) (0,1.2054431643) (0,1.2072685529) (0,1.2073734130) (0,1.2080997673) (0,1.2185471372) (0,1.2213885528) (0,1.2227011358) (0,1.2509629659) (0,1.2518364510) (0,1.2748260954) (0,1.2807316170) (0,1.2887474478) (0,1.2981197855) (0,1.3109543351) (0,1.3110622675) (0,1.3167599640) (0,1.3287112357) (0,1.3424644703) (0,1.3454749656) (0,1.3495100350) (0,1.3591733866) (0,1.3753837685) (0,1.3800020227) (0,1.3903707708) (0,1.3914588697) (0,1.3925653489) (0,1.3968589022) (0,1.4028022941) (0,1.4044553777) (0,1.4137126117) (0,1.4685240425) (0,1.4804672846) (0,1.4852072196) (0,1.5068487343) (0,1.5204673330) (0,1.5271685831) (0,1.6076900749) (0,1.6142708696) (0,1.6239794022) (0,1.6687339703) (0,1.7368576932) (0,1.7573486678) (0,1.7810107756) (0,1.7829729449) (0,1.8080983034) (0,1.8092252850) (0,1.8344241362) (0,1.8439146765) (0,1.8855156964) (0,1.9391689361) (0,2.0712787306) (0,2.1248530089) (0,2.2276325622) (0,2.2963481522) (0,2.6361137768)};
\node[anchor=south,font=\scriptsize,yshift=-7pt] at (axis cs:5,3.0315308433) {$n=2461$};
\end{axis}
\end{tikzpicture}

\captionof{figure}{
Time to first mutation detection for detected trials.
Boxes show the interquartile range and median.
Whiskers extend to the most extreme observations within
1.5 interquartile ranges of the box.
Labels report the number of detected trials out of 2500.
}
\label{fig:time_until_detection}

\end{minipage}
\hfill
\begin{minipage}[t]{0.49\textwidth}
\centering

\begin{tikzpicture}
\begin{axis}[
    width=\linewidth,
    height=5.5cm,
    boxplot/draw direction=y,
    boxplot/box extend=0.5,
    ymin=0,
    enlarge y limits={upper,value=0.15},
    xmin=0.5,
    xmax=5.5,
    ymax=110,
    ylabel={Test cases to first detection},
    xtick={1,2,3,4,5},
    xticklabels={
        Baseline,
        {LC\\only},
        {LC\\PERM},
        {All},
        {CNOT\\PERM}
    },
    xticklabel style={align=center,font=\scriptsize},
    tick align=outside,
    ymajorgrids=true,
    grid style={gray!20},
    clip=false
]
\addplot[boxplot prepared={draw position=1,lower whisker=1.0000000000,lower quartile=3.0000000000,median=6.0000000000,upper quartile=11.0000000000,upper whisker=23.0000000000},fill=boxbaseline,draw=black!70,mark=none] coordinates {(0,24.0000000000) (0,24.0000000000) (0,24.0000000000) (0,24.0000000000) (0,24.0000000000) (0,24.0000000000) (0,24.0000000000) (0,24.0000000000) (0,24.0000000000) (0,24.0000000000) (0,24.0000000000) (0,24.0000000000) (0,24.0000000000) (0,24.0000000000) (0,24.0000000000) (0,24.0000000000) (0,24.0000000000) (0,24.0000000000) (0,25.0000000000) (0,25.0000000000) (0,25.0000000000) (0,25.0000000000) (0,25.0000000000) (0,25.0000000000) (0,25.0000000000) (0,25.0000000000) (0,25.0000000000) (0,25.0000000000) (0,25.0000000000) (0,25.0000000000) (0,25.0000000000) (0,25.0000000000) (0,25.0000000000) (0,25.0000000000) (0,25.0000000000) (0,25.0000000000) (0,25.0000000000) (0,26.0000000000) (0,26.0000000000) (0,26.0000000000) (0,26.0000000000) (0,26.0000000000) (0,26.0000000000) (0,26.0000000000) (0,27.0000000000) (0,27.0000000000) (0,28.0000000000) (0,28.0000000000) (0,28.0000000000) (0,28.0000000000) (0,28.0000000000) (0,28.0000000000) (0,29.0000000000) (0,29.0000000000) (0,29.0000000000) (0,29.0000000000) (0,29.0000000000) (0,29.0000000000) (0,29.0000000000) (0,29.0000000000) (0,29.0000000000) (0,29.0000000000) (0,29.0000000000) (0,29.0000000000) (0,29.0000000000) (0,30.0000000000) (0,30.0000000000) (0,30.0000000000) (0,30.0000000000) (0,30.0000000000) (0,30.0000000000) (0,30.0000000000) (0,30.0000000000) (0,30.0000000000) (0,30.0000000000) (0,31.0000000000) (0,31.0000000000) (0,31.0000000000) (0,31.0000000000) (0,32.0000000000) (0,32.0000000000) (0,32.0000000000) (0,32.0000000000) (0,33.0000000000) (0,33.0000000000) (0,33.0000000000) (0,33.0000000000) (0,33.0000000000) (0,34.0000000000) (0,34.0000000000) (0,34.0000000000) (0,35.0000000000) (0,36.0000000000) (0,36.0000000000) (0,36.0000000000) (0,36.0000000000) (0,36.0000000000) (0,36.0000000000) (0,37.0000000000) (0,37.0000000000) (0,37.0000000000) (0,37.0000000000) (0,37.0000000000) (0,38.0000000000) (0,38.0000000000) (0,40.0000000000) (0,40.0000000000) (0,40.0000000000) (0,40.0000000000) (0,41.0000000000) (0,41.0000000000) (0,42.0000000000) (0,44.0000000000) (0,45.0000000000) (0,45.0000000000) (0,46.0000000000) (0,46.0000000000) (0,47.0000000000) (0,47.0000000000) (0,48.0000000000) (0,49.0000000000) (0,49.0000000000) (0,49.0000000000) (0,50.0000000000) (0,52.0000000000) (0,58.0000000000) (0,59.0000000000) (0,59.0000000000) (0,60.0000000000) (0,60.0000000000) (0,64.0000000000) (0,66.0000000000) (0,70.0000000000) (0,72.0000000000) (0,75.0000000000) (0,78.0000000000) (0,91.0000000000) (0,98.0000000000) (0,101.0000000000) (0,110.0000000000)};
\node[anchor=south,align=center,font=\scriptsize,yshift=5pt] at (axis cs:1,23.0000000000) {Mean:\\8.3};
\addplot[boxplot prepared={draw position=2,lower whisker=1.0000000000,lower quartile=2.0000000000,median=8.0000000000,upper quartile=30.0000000000,upper whisker=72.0000000000},fill=boxlconly,draw=black!70,mark=none] coordinates {(0,73.0000000000) (0,74.0000000000) (0,74.0000000000) (0,74.0000000000) (0,74.0000000000) (0,75.0000000000) (0,75.0000000000) (0,75.0000000000) (0,77.0000000000) (0,82.0000000000) (0,86.0000000000) (0,86.0000000000) (0,87.0000000000) (0,89.0000000000) (0,92.0000000000) (0,95.0000000000) (0,103.0000000000) (0,103.0000000000) (0,103.0000000000) (0,103.0000000000) (0,103.0000000000) (0,103.0000000000) (0,103.0000000000) (0,103.0000000000) (0,103.0000000000) (0,103.0000000000) (0,103.0000000000) (0,103.0000000000) (0,103.0000000000) (0,103.0000000000) (0,103.0000000000) (0,103.0000000000) (0,103.0000000000) (0,103.0000000000) (0,103.0000000000) (0,103.0000000000) (0,103.0000000000) (0,103.0000000000) (0,103.0000000000) (0,103.0000000000) (0,103.0000000000) (0,103.0000000000) (0,103.0000000000) (0,103.0000000000) (0,103.0000000000) (0,103.0000000000) (0,103.0000000000) (0,103.0000000000) (0,103.0000000000) (0,103.0000000000) (0,103.0000000000) (0,103.0000000000) (0,103.0000000000) (0,103.0000000000) (0,103.0000000000) (0,103.0000000000) (0,103.0000000000) (0,103.0000000000) (0,103.0000000000) (0,103.0000000000) (0,103.0000000000) (0,103.0000000000) (0,103.0000000000) (0,103.0000000000) (0,103.0000000000) (0,103.0000000000) (0,103.0000000000) (0,103.0000000000) (0,103.0000000000) (0,103.0000000000) (0,103.0000000000) (0,103.0000000000) (0,103.0000000000) (0,103.0000000000) (0,104.0000000000) (0,104.0000000000) (0,104.0000000000) (0,104.0000000000) (0,104.0000000000) (0,104.0000000000) (0,104.0000000000) (0,104.0000000000) (0,104.0000000000) (0,104.0000000000) (0,104.0000000000) (0,105.0000000000) (0,105.0000000000) (0,105.0000000000) (0,105.0000000000) (0,105.0000000000) (0,105.0000000000) (0,106.0000000000) (0,106.0000000000) (0,106.0000000000) (0,106.0000000000) (0,106.0000000000) (0,106.0000000000) (0,106.0000000000) (0,107.0000000000) (0,108.0000000000) (0,108.0000000000) (0,108.0000000000) (0,108.0000000000) (0,108.0000000000) (0,109.0000000000) (0,109.0000000000) (0,109.0000000000) (0,110.0000000000) (0,110.0000000000) (0,110.0000000000) (0,110.0000000000) (0,111.0000000000) (0,111.0000000000) (0,112.0000000000) (0,115.0000000000) (0,117.0000000000) (0,121.0000000000) (0,123.0000000000) (0,124.0000000000) (0,124.0000000000) (0,126.0000000000) (0,129.0000000000) (0,129.0000000000) (0,130.0000000000) (0,131.0000000000) (0,135.0000000000) (0,140.0000000000) (0,143.0000000000) (0,144.0000000000) (0,154.0000000000) (0,154.0000000000) (0,154.0000000000) (0,154.0000000000) (0,154.0000000000) (0,154.0000000000) (0,154.0000000000) (0,154.0000000000) (0,154.0000000000) (0,154.0000000000) (0,154.0000000000) (0,154.0000000000) (0,154.0000000000) (0,154.0000000000) (0,154.0000000000) (0,154.0000000000) (0,154.0000000000) (0,154.0000000000) (0,154.0000000000) (0,154.0000000000) (0,154.0000000000) (0,154.0000000000) (0,154.0000000000) (0,154.0000000000) (0,154.0000000000) (0,154.0000000000) (0,154.0000000000) (0,154.0000000000) (0,154.0000000000) (0,154.0000000000) (0,154.0000000000) (0,154.0000000000) (0,154.0000000000) (0,154.0000000000) (0,154.0000000000) (0,154.0000000000) (0,154.0000000000) (0,154.0000000000) (0,154.0000000000) (0,154.0000000000) (0,154.0000000000) (0,154.0000000000) (0,154.0000000000) (0,154.0000000000) (0,154.0000000000) (0,154.0000000000) (0,154.0000000000) (0,154.0000000000) (0,154.0000000000) (0,154.0000000000) (0,154.0000000000) (0,154.0000000000) (0,155.0000000000) (0,155.0000000000) (0,155.0000000000) (0,155.0000000000) (0,155.0000000000) (0,155.0000000000) (0,155.0000000000) (0,155.0000000000) (0,155.0000000000) (0,155.0000000000) (0,156.0000000000) (0,156.0000000000) (0,156.0000000000) (0,156.0000000000) (0,156.0000000000) (0,156.0000000000) (0,156.0000000000) (0,157.0000000000) (0,157.0000000000) (0,157.0000000000) (0,157.0000000000) (0,157.0000000000) (0,158.0000000000) (0,158.0000000000) (0,158.0000000000) (0,159.0000000000) (0,159.0000000000) (0,159.0000000000) (0,159.0000000000) (0,160.0000000000) (0,160.0000000000) (0,161.0000000000) (0,162.0000000000) (0,162.0000000000) (0,163.0000000000) (0,163.0000000000) (0,164.0000000000) (0,164.0000000000) (0,165.0000000000) (0,165.0000000000) (0,165.0000000000) (0,166.0000000000) (0,171.0000000000) (0,173.0000000000) (0,176.0000000000)};
\node[anchor=south,align=center,font=\scriptsize,yshift=5pt] at (axis cs:2,72.0000000000) {Mean:\\26.4};
\addplot[boxplot prepared={draw position=3,lower whisker=1.0000000000,lower quartile=3.0000000000,median=6.0000000000,upper quartile=12.0000000000,upper whisker=25.0000000000},fill=boxlcshuffle,draw=black!70,mark=none] coordinates {(0,26.0000000000) (0,26.0000000000) (0,26.0000000000) (0,26.0000000000) (0,26.0000000000) (0,26.0000000000) (0,26.0000000000) (0,27.0000000000) (0,27.0000000000) (0,27.0000000000) (0,27.0000000000) (0,27.0000000000) (0,27.0000000000) (0,27.0000000000) (0,27.0000000000) (0,27.0000000000) (0,27.0000000000) (0,27.0000000000) (0,27.0000000000) (0,27.0000000000) (0,27.0000000000) (0,27.0000000000) (0,27.0000000000) (0,27.0000000000) (0,28.0000000000) (0,28.0000000000) (0,28.0000000000) (0,28.0000000000) (0,28.0000000000) (0,28.0000000000) (0,28.0000000000) (0,28.0000000000) (0,29.0000000000) (0,29.0000000000) (0,29.0000000000) (0,29.0000000000) (0,29.0000000000) (0,29.0000000000) (0,29.0000000000) (0,30.0000000000) (0,30.0000000000) (0,30.0000000000) (0,30.0000000000) (0,30.0000000000) (0,30.0000000000) (0,30.0000000000) (0,30.0000000000) (0,30.0000000000) (0,30.0000000000) (0,30.0000000000) (0,30.0000000000) (0,30.0000000000) (0,31.0000000000) (0,31.0000000000) (0,31.0000000000) (0,31.0000000000) (0,31.0000000000) (0,31.0000000000) (0,31.0000000000) (0,32.0000000000) (0,32.0000000000) (0,32.0000000000) (0,32.0000000000) (0,32.0000000000) (0,32.0000000000) (0,32.0000000000) (0,32.0000000000) (0,32.0000000000) (0,32.0000000000) (0,32.0000000000) (0,33.0000000000) (0,33.0000000000) (0,33.0000000000) (0,33.0000000000) (0,33.0000000000) (0,33.0000000000) (0,33.0000000000) (0,33.0000000000) (0,33.0000000000) (0,33.0000000000) (0,33.0000000000) (0,33.0000000000) (0,33.0000000000) (0,34.0000000000) (0,34.0000000000) (0,34.0000000000) (0,34.0000000000) (0,34.0000000000) (0,34.0000000000) (0,34.0000000000) (0,35.0000000000) (0,35.0000000000) (0,35.0000000000) (0,36.0000000000) (0,36.0000000000) (0,37.0000000000) (0,37.0000000000) (0,37.0000000000) (0,37.0000000000) (0,37.0000000000) (0,37.0000000000) (0,37.0000000000) (0,37.0000000000) (0,37.0000000000) (0,38.0000000000) (0,39.0000000000) (0,39.0000000000) (0,39.0000000000) (0,39.0000000000) (0,39.0000000000) (0,39.0000000000) (0,40.0000000000) (0,40.0000000000) (0,40.0000000000) (0,41.0000000000) (0,41.0000000000) (0,41.0000000000) (0,41.0000000000) (0,41.0000000000) (0,41.0000000000) (0,41.0000000000) (0,42.0000000000) (0,42.0000000000) (0,43.0000000000) (0,43.0000000000) (0,43.0000000000) (0,43.0000000000) (0,44.0000000000) (0,44.0000000000) (0,44.0000000000) (0,45.0000000000) (0,45.0000000000) (0,45.0000000000) (0,45.0000000000) (0,45.0000000000) (0,46.0000000000) (0,46.0000000000) (0,46.0000000000) (0,46.0000000000) (0,46.0000000000) (0,47.0000000000) (0,48.0000000000) (0,51.0000000000) (0,52.0000000000) (0,52.0000000000) (0,55.0000000000) (0,55.0000000000) (0,56.0000000000) (0,56.0000000000) (0,59.0000000000) (0,59.0000000000) (0,61.0000000000) (0,61.0000000000) (0,63.0000000000) (0,64.0000000000) (0,67.0000000000) (0,74.0000000000) (0,77.0000000000) (0,83.0000000000) (0,93.0000000000) (0,102.0000000000) (0,180.0000000000)};
\node[anchor=south,align=center,font=\scriptsize,yshift=5pt] at (axis cs:3,25.0000000000) {Mean:\\9.4};
\addplot[boxplot prepared={draw position=4,lower whisker=1.0000000000,lower quartile=3.0000000000,median=6.0000000000,upper quartile=10.0000000000,upper whisker=20.0000000000},fill=boxall,draw=black!70,mark=none] coordinates {(0,21.0000000000) (0,21.0000000000) (0,21.0000000000) (0,21.0000000000) (0,21.0000000000) (0,21.0000000000) (0,21.0000000000) (0,21.0000000000) (0,21.0000000000) (0,21.0000000000) (0,21.0000000000) (0,21.0000000000) (0,21.0000000000) (0,21.0000000000) (0,21.0000000000) (0,21.0000000000) (0,22.0000000000) (0,22.0000000000) (0,22.0000000000) (0,22.0000000000) (0,22.0000000000) (0,22.0000000000) (0,22.0000000000) (0,22.0000000000) (0,22.0000000000) (0,22.0000000000) (0,22.0000000000) (0,23.0000000000) (0,23.0000000000) (0,23.0000000000) (0,23.0000000000) (0,23.0000000000) (0,23.0000000000) (0,23.0000000000) (0,24.0000000000) (0,24.0000000000) (0,24.0000000000) (0,24.0000000000) (0,24.0000000000) (0,24.0000000000) (0,24.0000000000) (0,24.0000000000) (0,24.0000000000) (0,24.0000000000) (0,24.0000000000) (0,24.0000000000) (0,24.0000000000) (0,24.0000000000) (0,24.0000000000) (0,25.0000000000) (0,25.0000000000) (0,25.0000000000) (0,25.0000000000) (0,25.0000000000) (0,26.0000000000) (0,26.0000000000) (0,26.0000000000) (0,26.0000000000) (0,26.0000000000) (0,26.0000000000) (0,26.0000000000) (0,26.0000000000) (0,27.0000000000) (0,27.0000000000) (0,27.0000000000) (0,27.0000000000) (0,27.0000000000) (0,27.0000000000) (0,27.0000000000) (0,28.0000000000) (0,28.0000000000) (0,28.0000000000) (0,28.0000000000) (0,28.0000000000) (0,28.0000000000) (0,28.0000000000) (0,29.0000000000) (0,29.0000000000) (0,30.0000000000) (0,30.0000000000) (0,30.0000000000) (0,31.0000000000) (0,31.0000000000) (0,31.0000000000) (0,32.0000000000) (0,32.0000000000) (0,32.0000000000) (0,32.0000000000) (0,33.0000000000) (0,34.0000000000) (0,34.0000000000) (0,36.0000000000) (0,36.0000000000) (0,37.0000000000) (0,37.0000000000) (0,37.0000000000) (0,39.0000000000) (0,45.0000000000) (0,47.0000000000) (0,64.0000000000) (0,94.0000000000) (0,140.0000000000)};
\node[anchor=south,align=center,font=\scriptsize,yshift=5pt] at (axis cs:4,20.0000000000) {Mean:\\7.4};
\addplot[boxplot prepared={draw position=5,lower whisker=1.0000000000,lower quartile=3.0000000000,median=5.0000000000,upper quartile=9.0000000000,upper whisker=18.0000000000},fill=boxcnot,draw=black!70,mark=none] coordinates {(0,19.0000000000) (0,19.0000000000) (0,19.0000000000) (0,19.0000000000) (0,19.0000000000) (0,19.0000000000) (0,19.0000000000) (0,19.0000000000) (0,19.0000000000) (0,19.0000000000) (0,19.0000000000) (0,19.0000000000) (0,20.0000000000) (0,20.0000000000) (0,20.0000000000) (0,20.0000000000) (0,20.0000000000) (0,20.0000000000) (0,21.0000000000) (0,21.0000000000) (0,21.0000000000) (0,21.0000000000) (0,21.0000000000) (0,21.0000000000) (0,21.0000000000) (0,21.0000000000) (0,21.0000000000) (0,21.0000000000) (0,21.0000000000) (0,21.0000000000) (0,21.0000000000) (0,21.0000000000) (0,21.0000000000) (0,21.0000000000) (0,21.0000000000) (0,21.0000000000) (0,21.0000000000) (0,22.0000000000) (0,22.0000000000) (0,22.0000000000) (0,22.0000000000) (0,22.0000000000) (0,22.0000000000) (0,22.0000000000) (0,22.0000000000) (0,22.0000000000) (0,22.0000000000) (0,23.0000000000) (0,23.0000000000) (0,23.0000000000) (0,23.0000000000) (0,23.0000000000) (0,23.0000000000) (0,23.0000000000) (0,23.0000000000) (0,24.0000000000) (0,24.0000000000) (0,24.0000000000) (0,25.0000000000) (0,26.0000000000) (0,26.0000000000) (0,26.0000000000) (0,26.0000000000) (0,27.0000000000) (0,28.0000000000) (0,28.0000000000) (0,28.0000000000) (0,28.0000000000) (0,28.0000000000) (0,29.0000000000) (0,29.0000000000) (0,29.0000000000) (0,30.0000000000) (0,30.0000000000) (0,30.0000000000) (0,32.0000000000) (0,33.0000000000) (0,36.0000000000) (0,37.0000000000) (0,37.0000000000) (0,37.0000000000) (0,38.0000000000) (0,38.0000000000) (0,46.0000000000) (0,51.0000000000) (0,90.0000000000) (0,116.0000000000)};
\node[anchor=south,align=center,font=\scriptsize,yshift=5pt] at (axis cs:5,18.0000000000) {Mean:\\6.6};
\end{axis}
\end{tikzpicture}

\captionof{figure}{
Number of test cases to first mutation detection for detected trials. Boxes show the interquartile range and median.
Whiskers extend to the most extreme observations within
1.5 interquartile ranges of the box.
Labels report the mean number of needed test cases.
}
\label{fig:test_cases_until_detection}

\end{minipage}

\end{figure*}

\textbf{RQ2a}. Figure \ref{fig:fault_detection_insert_gate} shows the fault detection rate per trial for the different methodologies under the \textit{insert random gate} mutation operator. The baseline and all metamorphic testing approaches except \textit{LC\_only} achieve similar detection rates, with 91.6\% for the baseline and approximately 92.0\% for the remaining metamorphic approaches. In contrast, \textit{LC\_only} achieves a lower detection rate of 70.4\%, corresponding to 352 detected mutations out of 500. None of the approaches reaches a 100\% detection rate. In total, 38 out of the 500 mutations were detected by none of the methods, while each of the remaining 462 was detected by at least one. We attribute these misses mainly to equivalent mutations, which do not change the prepared state and therefore cannot be detected by any method.
\begin{figure}[t]
\centering

\pgfplotsset{
    fault panel/.style={
        width=\linewidth,
        height=5cm,
        ybar,
        bar shift=0pt,
        bar width=20pt,
        ymin=35,
        ymax=115,
        xmin=0.5,
        xmax=5.5,
        ymax = 119,
        ylabel={Fault detection rate (\%)},
        ytick={0,20,40,60,80,100},
        xtick={1,2,3,4,5},
        xticklabels={
            Baseline,
            {LC\\only},
            {LC\\PERM},
            All,
            {CNOT\\PERM}
        },
        xticklabel style={align=center,font=\scriptsize},
        yticklabel style={font=\scriptsize},
        ylabel style={font=\small},
        tick align=outside,
        ymajorgrids=true,
        grid style={gray!20}
    }
}

\begin{minipage}[t]{0.49\linewidth}
    \centering
    \begin{tikzpicture}
    \begin{axis}[fault panel]
\addplot[fill=boxbaseline,draw=white] coordinates {(1,91.6000000000)};
\node[anchor=south,align=center,font=\tiny,yshift=10pt] at (axis cs:1,91.6000000000) {91.6\%\\(458/500)};
\addplot[fill=boxlconly,draw=white] coordinates {(2,70.4000000000)};
\node[anchor=south,align=center,font=\tiny,yshift=10pt] at (axis cs:2,70.4000000000) {70.4\%\\(352/500)};
\addplot[fill=boxlcshuffle,draw=white] coordinates {(3,92.0000000000)};
\node[anchor=south,align=center,font=\tiny,yshift=10pt] at (axis cs:3,92.0000000000) {92.0\%\\(460/500)};
\addplot[fill=boxall,draw=white] coordinates {(4,92.0000000000)};
\node[anchor=south,align=center,font=\tiny,yshift=10pt] at (axis cs:4,92.0000000000) {92.0\%\\(460/500)};
\addplot[fill=boxcnot,draw=white] coordinates {(5,92.2000000000)};
\node[anchor=south,align=center,font=\tiny,yshift=10pt] at (axis cs:5,92.2000000000) {92.2\%\\(461/500)};
    \end{axis}
    \end{tikzpicture}
    \caption{Fault detection rates for insert gate mutations.
    Labels show the detection percentage and detected trial count.}
    \label{fig:fault_detection_insert_gate}
\end{minipage}
\hfill
\begin{minipage}[t]{0.49\linewidth}
    \centering
    \begin{tikzpicture}
    \begin{axis}[fault panel]
\addplot[fill=boxbaseline,draw=white] coordinates {(1,100.0000000000)};
\node[anchor=south,align=center,font=\tiny,yshift=3pt] at (axis cs:1,100.0000000000) {100.0\%\\(500/500)};
\addplot[fill=boxlconly,draw=white] coordinates {(2,52.4000000000)};
\node[anchor=south,align=center,font=\tiny,yshift=3pt] at (axis cs:2,52.4000000000) {52.4\%\\(262/500)};
\addplot[fill=boxlcshuffle,draw=white] coordinates {(3,100.0000000000)};
\node[anchor=south,align=center,font=\tiny,yshift=3pt] at (axis cs:3,100.0000000000) {100.0\%\\(500/500)};
\addplot[fill=boxall,draw=white] coordinates {(4,100.0000000000)};
\node[anchor=south,align=center,font=\tiny,yshift=3pt] at (axis cs:4,100.0000000000) {100.0\%\\(500/500)};
\addplot[fill=boxcnot,draw=white] coordinates {(5,100.0000000000)};
\node[anchor=south,align=center,font=\tiny,yshift=3pt] at (axis cs:5,100.0000000000) {100.0\%\\(500/500)};
    \end{axis}
    \end{tikzpicture}
    \caption{Fault detection rates for incorrect qubit initialization
    mutations. Labels show the detection percentage and detected
    trial count.}
    \label{fig:fault_detection_initialize_qubits}
\end{minipage}

\end{figure}

\textbf{RQ2b}. All 1010~test cases passed the reference execution without
compilation. Of the compiled test cases, 878~passed, none violated a stabilizer,
and none raised a compiler error. The remaining 132~compilations (13.1\%) timed
out, all of them with the \texttt{lookahead} routing method
(\texttt{LookaheadSwap}). Of the 290~compilations that used \texttt{lookahead},
132 (45.5\%) timed out, whereas none of the 234~compilations with \texttt{basic}
and none of the 275~compilations with \texttt{sabre} routing did. The timeouts
were not tied to a particular setting: they occurred at all four optimization
levels, with all five basis gate sets, on all four restricted topologies, with
directed and undirected coupling maps, and with all three layout methods. They
were most frequent at optimization level~0 (64~timeouts, compared with 36, 18
and 14 at levels~1, 2 and~3). The remaining 211~compilations targeted the all-to-all topology, which
requires no routing; none of them timed out. Of the timeouts, 131 occurred on
follow-up test cases and one on a source test case.

Since a timeout may also indicate slow but correct routing, we analyzed
all timeouts with debug logging of the routing pass. In every
analyzed case, \texttt{LookaheadSwap} first routes some gates and then enters a
state in which each routing step inserts \texttt{search\_depth} SWAP gates on
the same edge of the coupling map. These SWAPs cancel each other, or leave a
single net SWAP that alternates the layout between two states. As a result, the
number of remaining gates stays constant while the output circuit and the
memory consumption keep growing. The pass thus never terminates, and it raises
no error. We minimized a failing input to an 8-qubit circuit with six CZ gates
on a tree-shaped coupling map at optimization level~0. For the same input,
\texttt{sabre} and \texttt{basic} routing finish within milliseconds. We
reported the bug to the Qiskit developers, and a contributor has prepared a fix~\cite{qiskit_issue_17127}. We found no earlier report of this behavior.

\subsection{Discussion}

The results reveal trade-offs between computational efficiency and fault detection effectiveness across our baseline and different metamorphic testing approaches.  

Regarding \textbf{RQ1}, the results indicate that the metamorphic testing approaches that incorporate CNOT-based MRs detect mutations with both lower runtime and fewer test cases. This is relevant for two reasons. First, a lower runtime directly reduces the overall testing effort by decreasing the time required to generate and execute test cases. Second, requiring fewer test cases reduces the number of program executions needed to detect a mutation, which is particularly important when executions are performed on quantum hardware, where computational resources and access time can be costly.

Metamorphic relations based on CNOT operations are generally more expensive to apply during testing. However, when the testing process terminates as soon as a mutation is detected, the approaches incorporating CNOT-based MRs achieve a lower overall runtime than the baseline method. This can be explained by the trade-off introduced by these relations. CNOT-based MRs require larger transformations of the graph and more substantial modifications and optimization of the corresponding measurement circuits, increasing the cost of generating individual test cases. At the same time, these larger transformations enable a broader exploration of the graph space than more local transformations such as LC, allowing faults to be exposed in fewer test cases.
\textit{PERM} relations are also beneficial because they permute the assignment of graph vertices to qubits. This increases the diversity of the qubits involved in the measured stabilizers and can therefore expose faults that would otherwise remain outside the support of low-weight stabilizers. In particular, because low weight stabilizers act only on a limited subset of qubits, repeatedly using the same qubit assignment may leave some parts of the program insufficiently exercised. \textit{PERM} helps mitigate this limitation by redistributing the measured structure across different qubits.

\begin{rqanswer}{Answer to RQ1 (Efficiency)}
Quantum Metamorphic Testing is substantially more efficient than the baseline in our experiments. All evaluated MR combinations detect mutations faster than the baseline. \textit{All} and \textit{CNOT\_PERM} also need fewer test cases, whereas \textit{LC\_only} needs considerably more. Although CNOT-based relations are more expensive per test case, their broader exploration of the graph space leads to earlier fault detection and lower overall testing cost.
\end{rqanswer}

With respect to \textbf{RQ2}, most evaluated approaches achieve perfect mutation detection for the majority of mutation operator classes. The main exception is the quantum metamorphic testing configuration using only LC-based MRs, which shows lower detection rates for the \textit{insert random gate} and \textit{incorrect qubit initialization} operators. These results are consistent with the more limited exploration provided by LC-only transformations. In contrast, incorporating CNOT- or PERM-based MRs broadens the explored test space and yields detection effectiveness comparable to the baseline. The presence of equivalent mutations does not affect the comparison between most QMT methods and the baseline.

Additionally, we identified a previously unknown bug in the Qiskit transpiler:
the \texttt{LookaheadSwap} routing pass can loop forever \cite{qiskit_issue_17127}, which affected 45.5\%
of the compilations using it. The bug was exposed by the implicit oracle that
compilation must terminate, not by a violated metamorphic relation. Although
131 of the 132~timeouts occurred on MR-generated follow-ups, a source test case
and the minimized plain CZ reproducer also trigger it.

\begin{rqanswer}{Answer to RQ2 (Testing Effectiveness)}
Our results show that quantum metamorphic testing can achieve fault detection effectiveness comparable to the baseline approach. Using only LC-based metamorphic relations is insufficient, as it explores a more restricted portion of the testing space. However, incorporating CNOT- or PERM-based relations broadens this exploration and allows the metamorphic testing configurations to reach fault detection rates comparable to the baseline. Our test campaign also revealed a previously unknown bug in the Qiskit transpiler: the \texttt{LookaheadSwap} routing pass does not terminate on certain inputs. The bug was exposed by compilation timeouts.
\end{rqanswer}

\subsection{Threats to Validity}

\textbf{Internal Validity:} One potential threat concerns the correctness of the measurement-circuit transformations used by the metamorphic testing approach, as our evaluation assumes that these transformations are implemented correctly. Although most of the transformations are relatively simple, implementation errors, particularly in the transformation associated with CNOT operations, could affect the observed fault-detection results. Additionally, bugs or unintended behavior in Python or Qiskit libraries could influence the results independently of the techniques being evaluated, beyond those that we aim to detect through this work. To mitigate this threat, we manually verified the transformations on small instances for which the expected results could be independently determined and inspected intermediate outputs throughout the testing pipeline. \\
A second threat arises from the randomized generation of mutants and input graphs. For a given mutation-operator class, random sampling may produce instances that are unusually similar to one another or that are not representative of the broader set of possible mutations. This could introduce variability into the measured effectiveness of the testing methods. We mitigate this threat by executing a large number of trials for each mutation-operator class and by evaluating multiple independently generated input graphs within each trial, thereby reducing the influence of individual random selections on the aggregate results.

\textbf{External Validity:} This evaluation is based on a particular set of mutation operators, input graphs, and compiler faults. Although these cases cover different types of faults and program transformations, they cannot capture the full range of defects that may occur in quantum software. The reported fault-detection results should therefore be interpreted with respect to the fault models and program instances considered in this study.

\section{Related Work}
\label{sec:related}

Quantum software testing has received growing attention in recent years
(see~\cite{ramalho_roadmap_2025} for an overview). Early work focused on
assertions. Huang and Martonosi~\cite{huangStatisticalAssertionsValidating2019} proposed
statistical assertions that check classical, superposition, and
entanglement properties from repeated measurements, and
Li et al.~\cite{li_proq_2020} proposed projection-based assertions that
require fewer measurements. Later work adapted classical testing techniques
to quantum programs, including property-based
testing~\cite{honarvar_pbt_2020}, input and output coverage
criteria~\cite{ali_quito_2021}, search-based test
generation~\cite{wang_qusbt_2021}, mutation
analysis~\cite{mendiluze_muskit_2021}, and metamorphic
testing~\cite{abreu_mt_oracle_2022}. Two recent approaches exploit the
choice of measurement basis. Oldfield et al.~\cite{oldfield_projective_2025}
reduce program specifications so that tests can use projective measurements
in mixed Hadamard bases, and Muqeet et al.~\cite{qops_2024} define test
cases as commuting Pauli strings that can be measured simultaneously. All
these approaches test quantum programs against their specifications. A
second line of work, discussed next, tests the compilers and platforms that
execute these programs.

Wang et al.~\cite{wang_qdiff_2021} proposed QDiff, which replaces gates with
equivalent gate sequences (e.g., a $Z$ gate with two phase gates $S$, since
$S^2 = Z$) and compares the output distributions of the resulting variants
on Qiskit, Cirq, and PyQuil. Paltenghi and
Pradel~\cite{paltenghi_morphq_2023} proposed MorphQ, the first metamorphic
testing approach for quantum computing platforms. MorphQ modifies randomly
generated Qiskit programs in ways that should not affect their output, for
example by reordering the qubits or by exporting the circuit to QASM and
importing it again. With MorphQ, the authors found 13 bugs in Qiskit.
Iwumbwe et al.~\cite{qutefuzz} proposed QuteFuzz, which generates programs
with nested branching statements that depend on measurement results. By
comparing results across compiler settings, the authors found 17 bugs in
Pytket, Qiskit, and Cirq. With QITE, Paltenghi and
Pradel~\cite{paltenghi_qite_2025} extended this line of work to multiple
platforms: QASM programs are imported into a platform, transformed, and
exported back to QASM, yielding programs that should all be equivalent. This
revealed 17 bugs in Qiskit, PennyLane, Pytket, and BQSKit. More recently,
Luo et al.~\cite{luo_qemi_2026} proposed QEMI, which removes dead code, that
is, operations that do not affect the final state, and found 12 bugs in
Qiskit, Q\#, and Cirq.

Other approaches target different parts of the quantum software stack.
Fuzz4All~\cite{xia_fuzz4all_2024} uses a large language model to generate
test programs, with Qiskit among its targets, and
FuzzQ~\cite{klimis_fuzzq_2025} tests quantum simulators with circuits
generated from a formal model of QASM~3.0, revealing eight bugs in Qiskit
and Cirq. Finally, most of these approaches build on classical compiler
testing (see~\cite{chen_compiler_testing_survey_2020} for a survey), such as
Csmith~\cite{yang_csmith_2011}, which compares the outputs of different
compilers on generated C programs, and EMI~\cite{le_emi_2014}, whose idea
QEMI transfers to quantum programs.

\section{Conclusion and Future Work}
We have introduced a QMT framework to determine the suitable measurements to test quantum state preparations. Our approach is based on exploiting the structure of the target states, and deriving the corresponding measurements to be performed with metamorphic relations. We implemented this approach on quantum programs aiming at the generation of stabilizer states, whose preparation is widely used to benchmark quantum hardware~\cite{cao_generation_2023,cai2026generalizedmermininequalitiesbenchmarking}. For testing such programs, standard testing approaches face two difficulties: First, although stabilizer states can be classically simulated efficiently, simulation neither verifies the implementation itself nor identifies which measurements can practically be performed. Second, finding such low-weight measurements for an arbitrary stabilizer state is provably computationally hard.

By using metamorphic relations, we overcome both. The proposed approaches reduce the runtime by an order of magnitude with respect to the baseline method, and combining diverse metamorphic relations retains a fault detection effectiveness comparable to the baseline. Our test campaign also revealed a real bug in Qiskit, namely the non-termination of \texttt{LookaheadSwap}. We reported this issue to the Qiskit developers, including a minimal reproducer \cite{qiskit_issue_17127}. At the time of writing the manuscript, the issue is open, and a contributor has prepared a fix.

Our work addresses the task of testing quantum software, taking into account the measurement limitations of any possible near-term quantum hardware. Therefore, while operating at the software level, it bridges theory and implementation of quantum software testing. In turn, a natural next step is to equip our method with noise-robustness tailored to specific hardware, such as trapped ions or superconducting qubits. It would also be timely to apply the proposed method to non-stabilizer states with specific structure, as well as to explore the converse direction: deriving the metamorphic relations to determine optimal states to test given measurements. Being of inherently general applicability and suitable to be equipped with experimental noise robustness, the proposed method aims at bridging the gap between software and hardware at testing quantum computations. 

\section{Data Availability and methodology statement}
A replication package is available at \url{https://doi.org/10.6084/m9.figshare.35005246}. It contains the implementation of the proposed metamorphic testing approach, scripts for the mutation and compiler testing experiment, and instructions for running the code.

All experiments were executed on a workstation equipped with an Intel Core i7-10700 CPU (2.90\,GHz, 8 cores, 16 threads) and 32\,GB of RAM, running Debian 13. The algorithms were implemented in Python version 3.13.5 using Qiskit version 2.5.2 and executed without GPU acceleration. The reported runtimes correspond to wall-clock execution times measured during the experimental runs.

We used Claude Opus~5.5 (Anthropic) to assist with code generation for the
experimental framework. All generated code was reviewed and tested by the
authors. The same model also assisted in literature search, generating the figures and in improving the text.

\begin{acks}
This research was funded by the Deutsche Forschungsgemeinschaft (DFG, German Research Foundation, project number 563437167) and the German Federal Ministry of Research, Technology and Space (Projects BeRyQC, Grant No.\ 01N124023D, and PraQTisch, Grant No.\ 01N126223D).
\end{acks}

\newpage

\bibliographystyle{ACM-Reference-Format}
\bibliography{sample-base}

\end{document}